\documentclass[aps,prb,twocolumn,showpacs,floatfix]{revtex4}

\usepackage{amsmath}
\usepackage{epsfig}

\begin{document}

\title{Theory of momentum resolved tunneling into a short quantum wire}
\author{Gregory A. Fiete$^{1,2}$, Jiang Qian$^1$, Yaroslav Tserkovnyak$^1$, and Bertrand I. Halperin$^1$}

\affiliation{$^1$Lyman Laboratory of Physics, Harvard University, Cambridge, MA 02138, USA\\$^2$Kavli Institute for Theoretical Physics, University of California, Santa Barbara, CA 93106, USA}

\begin{abstract}

Motivated by recent tunneling experiments in the parallel wire
geometry, we calculate results for momentum resolved tunneling into a
short one-dimensional wire, containing a small number of electrons. We
derive some general theorems about the momentum dependence, and we
carry out exact calculations for up to $N=4$ electrons in the final
state, for a system with screened Coulomb interactions that models the
situation of the experiments. We also investigate the limit of large
$N$ using a Luttinger-liquid type analysis. We consider the
low-density regime, where the system is close to the Wigner crystal
limit, and where the energy scale for spin excitations can be much
lower than for charge excitations, and we consider temperatures
intermediate between the relevant spin energies and charge
excitations, as well as temperatures below both energy scales.

\end{abstract}

\date{\today}
\pacs{73.21.-b,71.10.Pm,73.21.Hb,73.23.Hk}

\maketitle


\section{Introduction}
\label{sec:intro}

\begin{figure}
\includegraphics[width=1.0\linewidth,clip=]{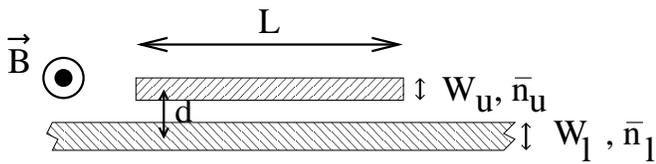}
\caption{\label{fig:geometry} Schematic geometry of electron tunneling
between two parallel quantum wires. Electrons are assumed to tunnel between an
infinitely long lower wire and a short upper wire of length $L$.  The wires are separated by a center-to-center distance $d$.  The upper (lower) wire has 
a width $W_u$ ($W_l$) and an average electron density $\bar n_u$ ($\bar n_l$).
A magnetic field $\vec B$ is applied perpendicular to the plane of the wires
to allow a field-dependent momentum boost $\hbar q_B=eBd$ in the tunneling. 
The energy of the tunneling electrons can be changed by adjusting the voltage
applied between the two wires.  In the experiments of 
Ref.~[\onlinecite{yacoby:sci02}], $d=30$ nm, $W_u$=20 nm, $W_l$=30 nm and 
$L=2-10\,\mu$m. Electron densities can range $\bar n=0-100\,\mu$m$^{-1}$. 
For much of this paper we will assume $(\bar n_l a_B)^{-1} \ll 1$ and 
$(\bar n_u a_B)^{-1}\gg1$, where $a_B$ is the Bohr radius,
so that the upper wire is close to  the Wigner crystal regime
and the lower wire can be treated as non-interacting.}
\end{figure}

Mesoscopic and nanoscale systems have proved to be fertile grounds for
studying quantum systems where interactions play a vital role in the
physics. In particular,
superconductivity,\cite{Black:prl96,Ralph:prl96} the Kondo
effect,\cite{Goldhaber:nat98,Manoharan:nat00,Nygard:nat00} and
Luttinger
Liquids,\cite{Bockrath:nat99,yacoby:sci02,yaro02,ishii:nat03,Auslaender:prl00}
have all been studied experimentally and theoretically to varying
degrees in these systems of reduced size and dimensionality.  Yet, due
to the myriad of energy scales that confinement and interactions
introduce, there is an ever unfolding landscape of rich physics to be
explored and understood.  Many of these interesting regimes in
mesoscopic and nanoscale systems can only be explored experimentally
with advances in technology.  One such advance was the ability to
fabricate quantum wires of extremely high quality in so-called cleaved
edge overgrowth samples.\cite{Pfeiffer:mej97} Auslaender {\it et
al.}\cite{yacoby:sci02} took this a step further to fabricate two {\em
parallel} quantum wires of very high quality which permitted momentum
to be approximately conserved in the tunneling between them.  See
Fig.~\ref{fig:geometry}.

Since then a series of experiments has been carried out at the
Weizmann Institute by Yacoby and
collaborators\cite{yacoby:sci02,yaro02,yaro03} on such cleaved edge
overgrowth samples fabricated on a GaAs substrate. The momentum
transfer in the tunneling was controlled by application of a magnetic
field perpendicular to the plane containing the two wires; the
energy-dependence was explored by varying the voltage between the wires; and
the electron density could be varied by application of voltage to
a top gate. With proper interpretation, these tunneling
measurements give very detailed information about the quantum wires.
For example, one can measure directly the Fermi wavevectors associated
with the occupied modes in the two wires; one sees diffraction fringes
due to the finite length of the shorter of the two wires, from which
one can deduce the shape of the gate potential that confines the
electrons at the ends of the shorter wire;\cite{Kakashvili:prl03} 
and one can see evidence of
the strong effects of electron-electron interactions in
one-dimensional systems, manifest in distinct propagation velocities
for spin and charge, as predicted by Luttinger-liquid
theory.\cite{yaro02,yaro03,Carpentier:prb02,Zulicke:prb02,Boese:prb01,Atland:prl99,Governale:prb00}
Earlier experiments on the {\em conductance} of finite quantum wires
motivated a number of theories in which the finite size effects were
explored in the context of Luttinger Liquid
theory.\cite{Eggert:prl96,Mattsson:prb97,Fabrizio:prb95,Kleimann:prb02,
Kleimann:prb00}

In a recent extension of the tunneling experiments, Steinberg {\em et
al.}\cite{Steinberg:sci04,Steinberg:sci05} have constructed a pair of wires in which a
central portion, of order $2\,\mu$m in length, is covered by a gate
and can be depleted of electrons relative to the rest of the wire.
Measurements revealed a dramatic transition, when the density of
electrons in the center of the {\em{upper}} wire was reduced below a
critical value.  The data give strong evidence that in this regime,
the electrons remaining in the depleted wire segment are separated
from the rest of the wire by barriers at the two ends, so that the
number $N$ of electrons in the wire segment becomes quantized in
integer values.  For example, the tunneling conductance showed a
series of sharp maxima, as the gate voltage was varied, analogous to
Coulomb blockade peaks, which were associated with transitions between
$N$ and $N-1$ electrons in the ground state of the wire segment.  The
positions of the Coulomb blockade peaks are independent of the applied
magnetic field, but there is variation of the peak height with field,
which enables one to measure the momentum distribution of the
many-body quantum state at the Fermi level. The momentum distribution
is quite different from what is observed at electron densities above
the critical density, and it was suggested that in the low-density
regime, the electrons are rather strongly localized by
electron-electron interactions as well as by the barriers at the end
of the wire segment.  It should be mentioned that this regime of
localized states was observed not only near depletion of the lowest mode
of the upper wire, but also as the electron density in the second and
third mode was reduced below a critical value.

At the present time, we do not have a good understanding of why a
potential barrier should arise spontaneously between the low density
central segment and the higher density ends of the upper wire.  Barriers of this type have been
found in Hartree-Fock calculations of a one-dimensional
model\cite{qian03} that is a crude representation of the geometry of
the experiments of Ref.~[\onlinecite{Steinberg:sci04,Steinberg:sci05}], as well as in
earlier LDA calculations of a quantum point contact.\cite{hirose03}
However, it remains to be seen whether these results are robust or are
artifacts of the particular models and approximations.

In any case, inspired by the results of Steinberg {\em et
al.},\cite{Steinberg:sci04,Steinberg:sci05} we have been led to consider the theory of
momentum-resolved tunneling into a short quantum wire, containing a
small number of electrons. We consider tunneling from an infinite
lower wire into a short upper wire, under conditions where the energy
levels of the upper wire are discrete, and the temperature is small,
at least compared to the Coulomb blockade energy and the Fermi energy
in the upper wire.  We will be particularly interested in the
situation where the density in the upper wire is small, so that when
two electrons come close together, the electron-electron repulsion is
strong compared to the average kinetic energy, and it is difficult for
two electrons to change places.  At low densities, it can be helpful
to think about the one-dimensional electron system as a kind of
fluctuating Wigner crystal,\cite{tanatar98,Glazman:prb92} with a
Heisenberg antiferromagnetic exchange
coupling\cite{Hausler:prb02,Matveev:prl04,Matveev:prb04} $J$ between
successive spins that is small compared to the energy for
short-wavelength charge fluctuations or the original Fermi energy.
Long-range interactions are expected to enhance the Wigner
crystal-like correlations,\cite{Schulz:prl93} and further reduce $J$.

For a finite low-density system, we can go one step further and
consider a situation where $J$ is so small that the temperature $T$
can be larger than $J/k_B$ but still small compared to the lowest energy
charge excitation, whose energy is proportional to the
charge-propagation velocity divided by the length of the wire.  This
situation, which we denote the ``free-spin regime,'' will be
considered in this paper, along with the strict low-temperature regime
(the ``ground-state tunneling regime'') where $k_BT$ is small compared to
all excitation energies. We shall also discuss a situation $k_BT \gg NJ$,
in which $k_BT$ is larger than all spin excitation energies, including
multiple spin excitations, which we shall refer to as the ``extreme
free-spin regime.''  We shall further distinguish between a ``true
Wigner crystal" and a ``fluctuating Wigner crystal" depending on
whether the root-mean-square displacement of an electron in the center
of a finite wire, due to quantum fluctuations, is smaller or larger
than the mean electron spacing.

In our discussion, we shall treat the infinite lower wire as a
collection of non-interacting electrons. See
Fig.~\ref{fig:geometry}. Interactions between the two wires are
considered only in so far as the lower wire may act as an ideal
conductor that contributes to a frequency-independent screening of the
Coulomb interaction between electrons in the upper wire.  The
high-density end-sections of the upper wire are taken into account
only as weakly coupled leads, which set the chemical potential in the
confined central part of the wire.

We shall concentrate on tunneling in the limit of zero bias voltage,
and shall examine both the momentum dependence (i.e., the dependence on
applied magnetic field) and the overall magnitude of the tunneling
conductivity, when the gate voltage is adjusted to give degeneracy
between the ground states of the upper wire for $N$ and $N-1$
particles.  Much of our discussion will focus on cases where $N$ is
quite small, but we shall also consider the behavior as $N$ becomes
large.

We assume that the coupling of the upper wire to its leads is weak
enough that the resulting level broadening is small compared to the
thermal energy $k_BT$.  However, we assume that the conductance between the
upper and lower wires is weaker still, so that the resistance is
dominated by the tunneling conductance, and so we can calculate the
current using Fermi's golden rule, in which the tunneling matrix
element enters only to second order, as a prefactor to the overall
conductivity.

This paper is organized in the following way.  In Sec.~\ref{sec:model}
we introduce the basic model and notation that we use throughout the
paper.  In Sec.~\ref{sec:noninteracting} we discuss the important
special case of non-interacting electrons in the upper wire.  In
Sec.~\ref{sec:theorems} we prove some general theorems about the
momentum structure of the tunneling for small $N$.  
In Sec.\ref{sec:smallN} we present a numerical study of the
double wire geometry, including interwire screening effects, for small
$N$ based on exact diagonalization.
In Sec.~\ref{sec:largeN} we discuss the general features expected for
tunneling in the limit of large $N$, focusing on the free-spin regime,
where the interactions are very strong and the temperature
is large compared to the spin exchange energies. In Sec.~\ref{sec:experiments}
we compare our theoretical findings with the measurements
by Steinberg {\em et al.},\cite{Steinberg:sci04,Steinberg:sci05}
before summarizing the paper in Sec.~\ref{sec:conclusions}.

\section{Model and notation}
\label{sec:model}

The Hamiltonian for electrons in the upper wire consists of a kinetic
energy, a one-body confining potential $U(x)$, and a two body
interaction $v(x-x^{\prime})$, where $x$ is the distance along the
wire. We shall discuss some theorems, below, which are independent of
the detailed form of $U$ and $v$.  In our more detailed calculations,
however, we shall concentrate on the case where $U$ contains infinite
vertical walls at $x=0 $ and $x=L$, together with a potential
arising from a uniform positive background in the interval $0 < x <
L$.  The interaction $v$ will have the form of a Coulomb potential
at intermediate distances, screened at long distances by the parallel
lower wire, and cut off at short distances because of the finite
thickness of the upper wire.  With these choices, one can consider a
limit $N \to \infty, L \to \infty$ with $N/L$ fixed, and the electrons
will be uniformly spread along the wire.  We shall
primarily be concerned, however, with systems where $N$ is not very
large.

Let $\Psi^N_\alpha$ and $\Psi^{N-1}_\gamma$ represent many body states of the upper
wire with $N$ and $N-1$ electrons, respectively, with energies $E^{N}_\alpha$ 
and $E^{N-1}_\gamma$.  We consider here the case where  the voltage $V$ between the upper and lower wire is small compared to $k_BT$, and  the tunneling conductance may be written as $G=G_++G_-$ in terms of
\begin{equation}
\label{eq:G}
G_\pm = C\mathcal{B}(k_\pm)\;,
\end{equation}
where
\begin{equation}
\label{eq:BT}
\mathcal{B}(k) = \sum_{\alpha \gamma \sigma} | \langle \Psi^N_\alpha | c^\dagger_{k \sigma} | \Psi^{N-1}_\gamma \rangle|^2 w_{\alpha \gamma}\; ,
\end{equation}
\begin{align}
\label{eq:wT}
w_{\alpha \gamma}&=e^{-\beta[E^{N-1}_\gamma - \mu (N-1)]} f(\epsilon_{\alpha \gamma}) \nonumber \\
&= e^{-\beta(E^N_\alpha - \mu N)} [1-f(\epsilon_{\alpha \gamma})]\;,
\end{align}
\begin{equation}
\epsilon_{\alpha \gamma} = E^{N}_\alpha - E^{N-1}_\gamma\;,
\end{equation}
\begin{equation}
C = \frac{\pi e^2}{2\hbar} \lambda^2 \beta \nu L \frac{e^{-\beta\mu N}}{Z_N + e^{-\beta\mu}Z_{N-1}}\;,
\label{eq:CT}
\end{equation}
and $c^\dagger_{k \sigma}=\int_0^Ldx\psi^\dagger_\sigma(x)e^{ikx}/\sqrt{L}$
creates a particle in the upper wire with
spatial wave function $e^{ikx}$ and spin $\sigma=\pm 1/2$, while
$\nu$ is the (total) density of states per unit length at the Fermi energy of the lower
wire, $\lambda$ is the coefficient of the tunneling Hamiltonian (properly normalized), $\mu$ is the chemical potential, $\beta=1/(k_BT)$,
$f(\epsilon)=1/\{1+\exp[\beta(\epsilon-\mu)]\}$ is the Fermi function, and $Z_N$ is the partition function of the upper wire when it contains precisely $N$ electrons.
(We neglect the probability of occupancies other than $N$ or $N-1$.)
$k_\pm$ are given by the lower wire wavevectors at the Fermi energy, shifted by the perpendicular magnetic field $B$:
\begin{equation}
k_\pm = \pm k_F^l+ e B d / \hbar\; .
\end{equation} 

In the limit where $k_BT$ is small compared to any excitation energies of the system, only the ground states contribute to the sum in (\ref{eq:BT}), and one obtains a large value of $G$ only if  the gate potential is adjusted so that $E^{N}-E^{N-1} - \mu$ is close to zero, i.e., of order $k_BT$ or smaller. If $E^N-E^{N-1}=\mu$ precisely, we have
\begin{equation}
G_\pm=\frac{\pi e^2}{4\hbar} \lambda^2 \beta \nu L \frac{1}{g_N + g_{N-1}}\mathcal{\tilde B}(k_\pm)\;,
\label{Gg}
\end{equation}
where
\begin{equation}
\mathcal{\tilde B}(k) = \sum_{\alpha \gamma \sigma} | \langle \Psi^N_\alpha | c^\dagger_{k \sigma} | \Psi^{N-1}_\gamma \rangle|^2\;,
\end{equation}
and 
 $g_N$ and $g_{N-1}$ are the degeneracies of the ground states $\Psi^N$ and
$\Psi^{N-1}$.

For the system under consideration, with spin-independent interactions
and no spin-orbit effects, eigenstates of the Hamiltonian will have a
definite quantum number for the total spin $S$.  In the absence of a
Zeeman field, then, the $N$-electron ground-state will have a
degeneracy $g_N = (2S_N+1)$, where $S_N$ is the corresponding spin
quantum number. The indices $\alpha$ and $\gamma$ may then be chosen
to label states with different values of the spin component $S_z$.  In
the case where there is an applied magnetic field with a Zeeman energy
large compared to $k_BT$, however, the degeneracy will be removed, and we
need only consider the states with spins aligned parallel to the
field, $S_z = S$.

When the Zeeman field is absent, we may use the 
Wigner-Eckart theorem to carry out the sum over the
spin indices, and write
\begin{equation}
\mathcal{\tilde B}(k) = D |M(k)|^2\;,
\end{equation}
where $M$ is the matrix element  
\begin{equation}
\label{eq:M}
M(k) =  \langle \Psi^N_\alpha | c^\dagger_{k \sigma} | \Psi^{N-1}_\gamma \rangle\;,
\end{equation}
with the indices for $S_z$ chosen as $\alpha=S_N$ and $ \gamma=S_{N-1}$, and
$ \sigma=\alpha-\gamma$.
The constant $D$ is given by
\begin{equation}
D=  1 + 2 \max(S_N,S_{N-1})\;.
\end{equation}
In order to have a non-zero matrix element, we must have
$|S_N-S_{N-1}| = 1/2.$ This condition is always satisfied for the
ground states, because for a system obeying the one-dimensional Schr\"odinger
Equation with spin-independent interactions, the lowest energy
state necessarily has spin quantum number $S=0$, if $N$ is even, and
$S=1/2$, if $ N$ is odd, according to the Lieb-Mattis
theorem.\cite{lieb62} This theorem also tells us that $D=2$ for
tunneling between ground states.\cite{Lieb:footnote} 


It is instructive to define a ``quasi-wavefunction''   
\begin{equation}
\Psi^{N}_{\rm eff}(x) \equiv \langle \Psi^{N-1}_\gamma|\psi_\sigma(x) |\Psi^{N}_\alpha \rangle \;,
 \label{qwf}
 \end{equation}
 so that $M(k) =\int dx e^{ikx} \Psi^{N*}_{\rm eff}(x)/\sqrt{L}$.
 For non-interacting electrons, $\Psi^{N}_{\rm eff}(x)$ is the wave function for the last added electron, and its norm is 1.  For interacting electrons, the norm will be less than 1,   
due to orthogonality catastrophe-type effects. 

We remark that the value of $G$ given by equation (\ref{Gg}) is not
actually the maximum value of the conductance that can be achieved
at the given temperature $T$.  A somewhat larger conductance can be
achieved by shifting the gate voltage slightly away from the value
where the states for $N$ and $N-1$ particles have precisely the same
energies, in order to take advantage of the differing degeneracies of
the states.  One finds that the maximum conductance is related to $G$ in Eq.~(\ref{Gg})
by
\begin{equation}
G_{\rm max}/ G = 2 (\eta^2 + 1) / (1 + \eta)^2\;, 
\end{equation}
where $\eta = (g_N / g_{N-1})^{1/2}$. The maximum occurs when  
$E^N - E^{N-1}-\mu = k_BT \ln \eta$, so that there is equal probability for the upper wire to have $N$ or  $N-1$ electrons.
For ground-state tunneling between states with spins 0 and 1/2, 
the correction factor for the maximum conductance is 
$G_{\rm max}/ G = 1.0294$.

If the spin excitation energies are sufficiently low compared to all
other excitation energies, we may consider tunneling in the ``extreme
free-spin'' regime, where $k_BT$ is large compared to the spin
energies, and we may neglect the energy differences between different
spin states.  The degeneracy factors in the expression for
$C$, Eq.~(\ref{eq:CT}), are then given by $g_N=2^N$, and the labels
$\alpha$ and $\gamma$ in the equation for $\mathcal{\tilde B}(k)$,
Eq.~(\ref{eq:BT}), must be summed over all of the states in the
manifold.

In the formulas above, we have assumed that the localized section of the upper wire is separated from its high-density leads by barriers sufficiently high that 
the level width $\Gamma$ due to coupling to the leads
is small compared to $k_BT$.      In the other limit, where $k_BT \ll \Gamma$,  the  prefactor $C$  for ground-state tunneling will be
proportional to $\Gamma^{-1}$ rather than $T^{-1}$.  If $eV$ is larger
than both $k_BT$ and $\Gamma$, then we cannot use the linear conductance formulas,
but the dependence of the current on momentum transfer should still be proportional to $\mathcal{\tilde B}(k)$ at moderate voltages.

\section{Non-interacting electrons}
\label{sec:noninteracting}

It is useful to begin by recalling some results for non-interacting
electrons.  We consider particularly the case of square-well
confinement, where $U(x)=0$ for $0< x < L$, and $\infty$ otherwise.
The one-electron states are then sine waves, with spatial dependence
$\phi_N(x) = (2/L)^{1/2} \sin [k_{\tilde{N}}x]$, where 
$k_{\tilde{N}} = \tilde{N} \pi / L$ , and, in the absence of a Zeeman field,
\begin{equation}
\tilde{N} = \left[ \frac{N+1}{2}\right]_{\rm Int}\;,
\end{equation}
with $[X]_{\rm Int}$  the integer part of $X$. For fully polarized electrons (or spinless fermions), we have simply $\tilde{N} = N$.   
In either case, one finds that 
\begin{align}
\label{eq:M1}
M(k)=&e^{i k L/2}\frac {\sqrt{2}i^{\tilde{N}-1}} {L} \Biggl[ \frac {\sin[(k+k_{\tilde{N}})L/2]}{k+k_{\tilde{N}}} 
\nonumber \\ &-(-1)^{\tilde{N}}
 \frac {\sin[(k-k_{\tilde{N}})L/2]} {k-k_{\tilde{N}}}            \Biggr]\;.
\end{align}

We note that the matrix element vanishes at $k=0$ if and only if $\tilde{N}$ is even.
\begin{figure}[ht]
\includegraphics[width=1.0\linewidth,clip=]{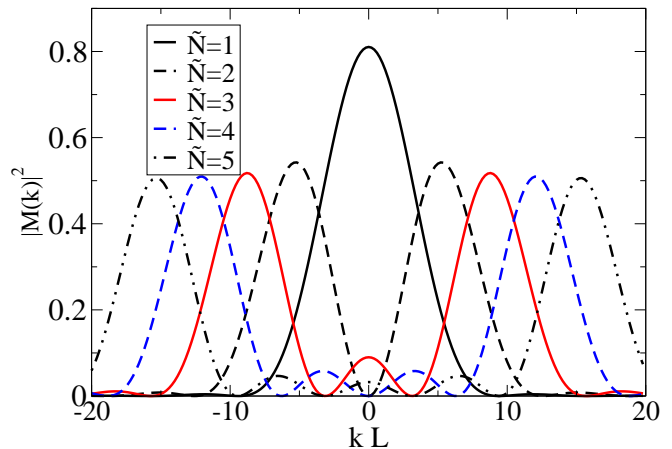}
\caption{\label{fig:FT01_box} (Color online.) Momentum resolution of one-particle box states.  Shown is the modulus squared of Eq.~(\ref{eq:M1}) for various values of $\tilde N$.
Many of the features in the non-interacting case remain when interactions are included and the Fourier
transform of a many-body wavefunction is taken.}
\end{figure} 

For large $N$, we see in Fig.~\ref{fig:FT01_box} that $|M(k)|^2$ has
sharp maxima at $k= \pm k_{\tilde{N}}$, which become $\delta$-functions in the
limit $N \to \infty, \; L \to \infty$.  Of course, $k_{\tilde{N}}$ corresponds to
the Fermi momentum of the infinite system. For large but finite $N$,
the $\delta$-functions broaden into diffraction patterns, with spacing
$\delta k = 2 \pi / L$ between successive zeroes, and a peak-width
of the same order.

Tserkovnyak {\em et al.}\cite{yaro02,yaro03} have discussed the form
of $M(k)$ for large but finite $N$ when the confinement at the end of
the wires is soft on the scale of the Fermi wavelength. In this case
they find an asymmetric diffraction pattern, which falls off rapidly
for $|k| > k_{\tilde N}$, but relatively slowly on the side $|k| < k_{\tilde N}$.  The
spacing between zeroes is not constant and is slightly larger than $2
\pi /L$.

\section{Some theorems for small systems}
\label{sec:theorems}

Some rigorous theorems are helpful for understanding the properties of
$\mathcal{\tilde B}(k)$, particularly at small values of $N$.

As a consequence of time-reversal invariance, all wave functions may
be chosen to be real.  From this it follows that $\mathcal{\tilde B}(k) = \mathcal{\tilde B}(-k)$.
Also, the matrix element $M(k)$, which determines $\mathcal{\tilde B}(k)$ in the case
of ground state tunneling, satisfies $M(k) = M(-k)^\ast$.

If the confining potential is symmetric about the center of the well, the
states $\Psi^N$ and $\Psi^{N-1}$ may be classified by their parity under
inversion through the center of the well. If the states have 
the same parity, then
$M(k) = M(-k)$, and hence $M$ must be real.  If the states have
opposite parity, then $M(k)=-M(-k)$, and $M$ must be purely
imaginary. In either case, if $M(k)$ changes sign, as a function of
$k$, there will be one or more values of $k$ where $M(k)=0$.  Thus, in
the ground-state tunneling regime, there can be points $k$ where
$\mathcal{\tilde B}(k)=0$.  We have already seen an example in the case of
non-interacting electrons.

For the case $N=1$, the ground state wave function can be written in
first quantized notation (in terms of position $x$ and spin index $\sigma$), as
\begin{equation}
\Psi_\alpha(x, \sigma) = \Psi(x) \delta_{\sigma, \alpha}\;,
\end{equation}
where $\Psi (x) > 0 $ for all $x$.  (This is true regardless of
whether or not the confining potential has inversion symmetry.)  Thus,
for the case $N=1$, the matrix element $M$ can be written as
\begin{equation}
M(k) =  \int  dx\frac{e^{ikx}}{\sqrt{L}} \Psi^* (x) \;.
\end{equation}
Clearly, $|M(k)|$ has its maximum when $k=0$, for $N=1$.

We obtain a similar behavior for ground-state tunneling in the case
$N=2$.  The ground state of the two-fermion problem, in any number of
dimensions, is always a spin-singlet, with a spatial wave function
that has no nodes.  Thus, for $N=2$ we may write, in first-quantized
notation,
\begin{equation}
|\Phi \rangle = f(x_1, x_2) \chi_0 (\sigma_1,\sigma_2)\;,
\end{equation}
where $\chi_0 = 2^{-1/2} ( | \uparrow \downarrow \rangle - |\downarrow
\uparrow \rangle) $, while $f$ is real and positive, and symmetric under
interchange of its arguments.  We may also write
\begin{equation}
 c^\dagger_{k \downarrow} | \Psi_\uparrow\rangle = \frac{1}{\sqrt{2L}} 
[e^{ikx_2} \Psi (x_1) \  | \uparrow \downarrow \rangle 
-e^{ikx_1} \Psi (x_2) \  | \downarrow \uparrow \rangle]\; ,  
\end{equation}
where $\Psi (x) > 0$ is the wave function for the one-particle
system. The matrix element $M(k)$ is then given by
\begin{equation}
M(k) = \int dx_1 dx_2 \frac{e^{ikx_2}}{\sqrt{L}} \Psi (x_1) f(x_1,x_2) \;,
\end{equation}
which clearly has its maximum at $k=0$.

The situation is different if we consider tunneling between the $N=1$
system and the lowest triplet state of the $N=2$ system. If we choose
the spins to be maximally aligned in the $z$-direction, we have
\begin{equation}
|\Phi\rangle = g(x_1,x_2) |\uparrow \uparrow\rangle ,
\end{equation}
and
\begin{equation}
c^\dagger_{k \uparrow} | \Psi_\uparrow\rangle = \frac{1}{\sqrt{2L}}
[e^{ikx_2} \Psi (x_1)  
-e^{ikx_1} \Psi (x_2) ] \  |\uparrow \uparrow \rangle \;, 
\end{equation}
where $g$ is antisymmetric under interchange of its arguments, and
\begin{equation}
M(k) = \sqrt{2} \int dx_1 dx_2 \frac{e^{ ikx_2}}{\sqrt{L}} \Psi (x_1) g(x_1,x_2) \;.
\end{equation}

If the confining potential is symmetric, then $g$ will be
antisymmetric under inversion, while $\Psi$ is symmetric.  It follows
that $M=0$ at $k=0$, and $|M(k)|$ will have its maximum at a non-zero
value of $k$.

The triplet state will be the actual ground state of the $N=2$ system
if there is an applied Zeeman field large enough to overcome the
exchange splitting between the singlet and triplet states. It is also
the correct ground state for a model of spinless fermions.  In the
absence of a Zeeman field, in the free-spin regime, where the
temperature is larger than the exchange splitting, but smaller than
all other excitation scales, the quantity $\mathcal{\tilde B}(k)$ will have two
contributions from the singlet state and four contributions from the
triplet.  In this case, the function $\mathcal{\tilde B}(k)$ will not be zero at $k=0$, but we may have a situation with $\mathcal{\tilde B}(k)$ having a local minimum at $k=0$, with a pair of maxima at a finite value of $|k|$, as in Fig.~\ref{fig:FT12} below.

For $N=3$, the many-body ground state for non-interacting electrons
in symmetric confinement
has odd parity under inversion, as it has two electrons in the lowest
even-parity level, and one electron in the second odd-parity
level. The $N=4$ state has even parity.  These parities cannot be
altered by a weak electron-electron interaction, and it is likely that
they will hold even for rather strong interactions.  As we have seen,
the ground state for $N=2$ has even parity. Consequently, for
electrons in a symmetric confining potential, the ground state
tunneling form factor $\mathcal{\tilde B}(k)$ should vanish at $k=0$ for $N=3$ and
$N=4$ .

The feature of a zero amplitude at $k=0$ does not persist, if the
confining potential is not symmetric under inversion.  For a
sufficiently asymmetric potential there need not be even a local
minimum near $k=0$.  For example in the case of an asymmetric
double-well potential, when there is little overlap between the wave
functions in the two wells, one would expect that $\mathcal{\tilde B}(k)$ for $N=3$ and
$N=4$ should be qualitatively similar to the form for $N=1$ and $N=2$.
 
The state $N=5$ has even parity, at least for weakly interacting
electrons.  Thus the matrix element for tunneling between the ground
states with 4 and 5 electrons will be non-zero at $k=0$.  For
non-interacting electrons, $\mathcal{\tilde B}(k)$ has a local maximum at $k=0$, but
has its absolute maxima at a non-zero value of $|k|$, with zeroes of
the amplitude in between. See Fig.~\ref{fig:FT01_box}. We
expect these qualitative features for non-interacting electrons to persist for strong interactions in small-$N$ wires.

We now turn to exact numerics for small $N$.

\section{Calculations for $N=1$ to $4$}
\label{sec:smallN}

We can illustrate some of the ideas and results discussed in
Sec.~\ref{sec:theorems} with exact numerical calculations for small
$N$ which take into account finite size and screening effects from the
lower wire.

\subsection{Form of electron interaction}

We assume the interaction potential is a Coulomb interaction with both
a short and a long range cutoff. The short range cutoff is achieved in
a simple way by modifying a $\frac{1}{z}$ potential to
$\frac{1}{\sqrt{z^{2}+W^{2}}}$ where $z=x-x'$ is the separation of two
electrons along the wire and $W$ is a short range cutoff ($W_u$ for
the upper wire), which is roughly the width of the quantum wire.  The
long range cutoff, with experiments by Steinberg \textit{et
al.}\cite{Steinberg:sci04,Steinberg:sci05} in mind, is achieved by putting a second,
lower wire with width $W_l$, assumed to be \emph{perfectly
conducting}, parallel to the system under study, at center to center
distance $d$.  See Fig.~\ref{fig:geometry}.  This approximation is
valid, strictly speaking, if the electron density in the lower wire is
sufficiently high that many modes are occupied, while there is only
one mode occupied in the central region of the upper wire. We note
that for the experiments of Ref.~[\onlinecite{Steinberg:sci04,Steinberg:sci05}], there
are at most only a few modes occupied in the higher density wire, so
the assumptions made here may overestimate the quality of screening
actually obtained in the experiments.

\begin{figure}
\includegraphics[width=1.0\linewidth,clip=]{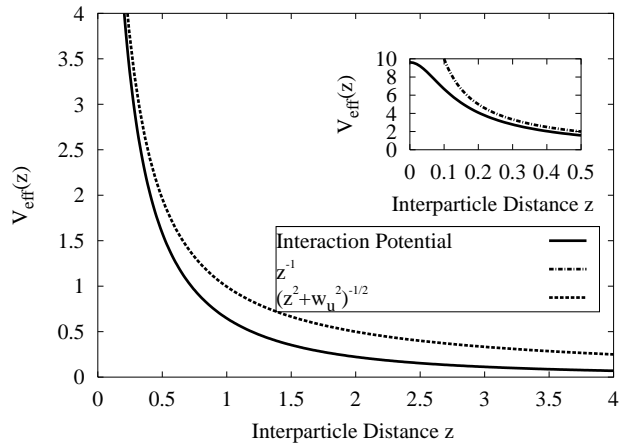}
\caption{\label{fig:potent} Effective interaction potential $V_{\rm eff}$, as a
    function of the separation $z$ between two electrons in the upper wire,
    taking into account the finite width $W_u$ of the upper wire and 
screening due to
a parallel lower wire of width $W_l$ at a distance $d$. See 
Fig.~\ref{fig:geometry}. Dashed and dotted lines show the
bare Coulomb interaction $z^{-1}$, and the unscreened interaction softened
at short distances, $1/\sqrt{z^2+W_u^2}$. Parameters are $d=1$, $W_u=0.1$ and
$W_l=0.15$. Region $0<z<0.5$ is expanded in the inset. In the main plot $1/z$ and $1/\sqrt{z^2+W_u^2}$ are indistinguishable in the range
$0.5<z<4$. In the inset the interaction potential $V_{\rm eff}(z)$ and  $1/\sqrt{z^2+W_u^2}$ are indistinguishable in the range $0<z<0.5$.}
\end{figure}

Let $V_{0}(z,W)=\frac{e^2}{\epsilon}\frac{1}{\sqrt{z^2+W^2}}$ be the
interaction with short range cutoff $W$ between particles
separated by a distance
$z$. Here $\epsilon$ is the dielectric constant. The energy
distribution of the wires in Fourier space can be estimated as:
\begin{align} \label{eq:eq1}
E=&\frac{1}{2}\int dq\biggl[\delta\rho_{u}(q)\delta\rho_{u}(q)\tilde V_0(q,W_u)+\delta\rho_{l}(q)\delta\rho_{l}(q)\tilde V_0(q,W_l)\nonumber\\
&+
                2\delta\rho_{u}(q)\delta\rho_{l}(q)\tilde V_0(q,d)\biggr]\;,
\end{align}
where $\delta\rho_{u/l}(q)$ is the net charge density (electron density 
minus background positive charge) of the upper/lower wire in Fourier space 
and $\tilde V_0(q,W_u/W_l)$ is the Fourier transform of the interaction (already with 
the short range cutoff) within the upper/lower wire. Taking the variation
with respect to the lower wire charge density $\delta\rho_{l}$:
\begin{equation}
\frac{dE}{d(\delta\rho_l)}=\delta\rho_{l}(q)\tilde V_0(q,W_l)
+\delta\rho_{u}(q)\tilde V_0(q,d)\;.
\end{equation}
Setting it to zero,
\begin{equation}
\delta\rho_{l}(q)=-\frac{\delta\rho_{u}(q)\tilde V_0(q,d)}{\tilde V_0(q,W_u)}\;.
\end{equation}
Putting this back into (\ref{eq:eq1}) and taking out the common factors of 
$\delta\rho_{u}^{2}$,
\begin{equation}
E=\frac{1}{2}\int\delta\rho_{u}^{2}(q)\left[\tilde V_0(q,W_u)-\frac{\tilde V_0^2(q,d)}{\tilde V_0(q,W_l)}\right]dq\;.
\end{equation}
Comparing this with the form of self interaction in the 
upper wire it is easy to see
the effective form of interaction in Fourier space is
\begin{equation}
\tilde V_{\rm eff}(q)=\tilde V_0(q, W_u)-\frac{\tilde V_0^2(q,d)}{\tilde V_0(q,W_l)}\;,
\label{eq:int}
\end{equation}
where we have made explicit reference to the short distance cut off in 
real space: $\tilde V_0(q,W)=\int_{-\infty}^{\infty}dx \frac{e^{iqx}}{\sqrt{x^2+W^2}}$.
The interaction (\ref{eq:int}) is plotted in real space in 
Fig.\ref{fig:potent}.

\subsection{Model for confinement and background charge distribution}

We now consider the model of a few interacting electrons in a box of
length $L$. A uniform positive background with density equal to the
average electron density is added to maintain charge neutrality. The
resulting one-body potential $U(x)$ takes the following form:
\begin{equation}
U(x) = \left 
	\{ \begin{array}{ll}
    \infty , & \textrm{if $x<0$ or $x>L$},\\ 
	-\bar n\int_{0}^{L}V_{\rm eff}(x-x')dx' , & \textrm{if $0<x<L$}.
	    \end{array}\right.
\end{equation}

Here $\bar n$ is the density of positive background charge. When the electron
number $N$ is fixed, $\bar n=N/L$. When the electron density 
changes, as in the case of tunneling into a relatively rigid box, $\bar
n$ is taken to be the average of the initial and final electron
density. The precise value of $\bar n$ does not qualitatively change the
results discussed below.

Systems with up to four electrons, both with and without spin, were
solved by directly diagonalizing the exact many-body Hamiltonian using
the Lanczos method.\cite{Dagotto:rmp94,lanczos} Earlier applications
of the Lanczos method to one-dimensional quantum dots were reported in
Refs.~[\onlinecite{Hausler:prb93,Jauregui:prb96,Jauregui:epl96}] , by
H\"ausler and Kramer, and collaborators. Energy spectra for up to
$N=4$ particles, with Coulomb interactions unscreened at large
distances, were presented in Ref.~[\onlinecite{Jauregui:epl96}], and
the splitting between the lowest energy singlet and triplet states was
found to fall off rapidly with increasing wire length in this case.
Density distributions for up to four particles, and density
correlation functions for three particles, were obtained in
Ref.~[\onlinecite{Jauregui:epl96}]. Matrix elements for tunneling were
discussed, and tunneling rates were computed, in
Ref~[\onlinecite{Jauregui:prb96}], but only for the incoherent case,
with no momentum conservation. The parameters of the
calculation, including the interaction strength and the geometry of
two parallel wires with one screening the other, were obtained from
estimates in the experiments by Steinberg \textit{et
al.}.\cite{Steinberg:sci04,Steinberg:sci05} In the calculation shown
here we take the upper wire width $W_u$ to be 20 nm, the lower wire
width $W_l$ to be 30 nm and the center to center distance $d$ between
wires to be 31 nm. The interaction strength between electrons is
obtained by using the dielectric constant of GaAs, $\varepsilon=13.1$,
and effective mass $m^{*}=0.067m_{\rm{e}}$, which gives a Bohr radius
$a_B=\varepsilon \hbar^2/(m^*e^2)\approx 10$ nm.

\subsection{Ground state density and density-density correlation}

\begin{figure}
\includegraphics[width=1.0\linewidth,clip=]{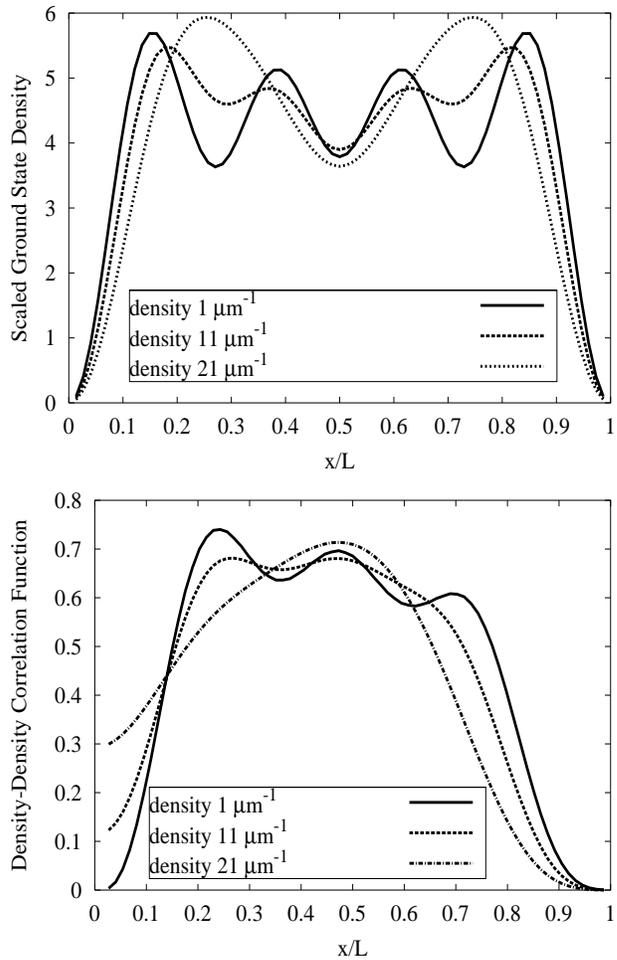}
\caption{\label{fig:density_spin} Top: Ground state density of a 
four electron system
    with spin.  Scaled densities are measured in units of $1/L$. 
	In the label of
	the individual plot, ``density'' means the \emph{average
	physical} density $N/L$ of the system, in units of $\mu\rm{m}^{-1}$. 
Bottom: Ground state density-density correlation
    of a four electron system with spin.
    The density-density correlation
	function is defined as 
	$\frac{1}{1-x}\int_{0}^{1-x}\rho (x')\rho (x'+x)dx'$. 
	The pre-factor before the integral is to take into account the finite 
	length of the system.}
\end{figure}

Before discussing the momentum structure of the tunneling matrix
element, a discussion of ground state properties is useful. In the
case of electrons with spin, as discussed before, the ground state has
total spin $0$, for a system with an even number of electrons, and
total spin $1/2$, for a system with an odd number of electrons. As the
electron density decreases, there is a relative increase in the
importance of the potential energy compared to the kinetic energy, and
there is a crossover from a relatively uniform liquid state into
a Wigner crystal-like state with spatially localized electrons. This
crossover can be seen in Figs.~\ref{fig:density_spin} and
\ref{fig:density_spin} where we plot the ground state density
distribution and the density-density correlation function for a system
of $N=4$ electrons.  

In Fig.~\ref{fig:density_spin}, at a large physical density of
twenty-one electrons per micron, the scaled electron density
distribution has smooth variation with spatial frequency $2k_{F}$,
which is simply a Friedel oscillation resulting from boundary
effects. At the lowest density of one electron per micron, the scaled
electron density has distinct peaks, whose number equals that of the
electrons in the system, corresponding to a $4k_{F}$ Wigner crystal
density variation. The signature of Wigner crystal correlations is
also clearly seen in Fig.~\ref{fig:density_spin}, in the form of
$4k_F$ oscillations in the density-density correlation function at low
physical density.

\begin{figure}
\includegraphics[width=1.0\linewidth,clip=]{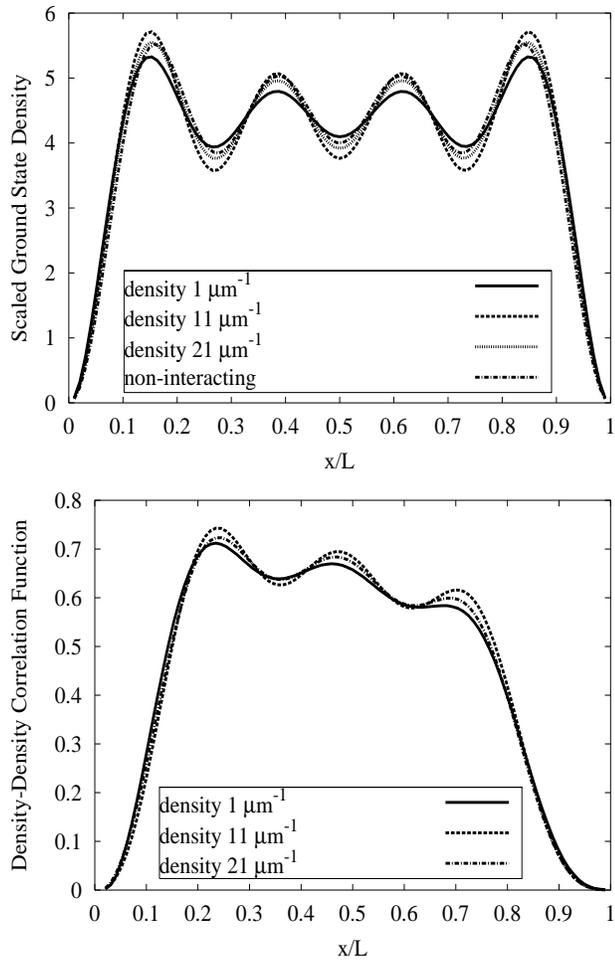}
\caption{\label{fig:density_spinless} Top: Ground state density of a 
four electron system without spin. Bottom: Ground state density-density 
correlation of a four electron system without spin. See Fig.~\ref{fig:density_spin} for a definition of the density-density correlation function.}
\end{figure}

The behavior of spinless electrons and spin polarized electrons is
significantly different from the case with spin (shown in 
Fig.~\ref{fig:density_spin}) in its spin-singlet ground state.  As shown in 
Fig.~\ref{fig:density_spinless}, contrary to the case of a
system with spin, decreasing the physical density does not
significantly alter the ground state. 

Neither the scaled electron
density distribution nor the density-density correlation function
changes more than ten percent when the physical density changes over
an order of magnitude from twenty-one to one electron per micron. This
means that the spinless system is Wigner crystal-like, although essentially non-interacting for all densities
and gate potentials. This is the result of a combination of the Pauli
Exclusion Principle, the cutoff of the interaction at large distances due
to the second perfectly conducting (lower) wire, and the actual parameters of
the experimental system. When the density is high and inter-particle
distance $z\ll a_B$, the interaction energy is much smaller
than kinetic energy. When the density is low and average
inter-particle distance $z\gg d$, the interaction becomes effectively
short-ranged due to screening, diminishing its effect because in the
spinless model the wavefunction vanishes when electrons come together. 

As mentioned before, in the experiments of Steinberg \textit{et
al.},\cite{Steinberg:sci04,Steinberg:sci05} $a_B\approx 10$ nm and
$d\approx 30$ nm. Define $r_{s}\equiv\ a/a_B$, where
$a$ is the average inter-particle distance. Then, for
$r_{s}\gtrsim 3$, the long range cutoff set by screening from the second
wire kicks in and starts to strongly suppress interaction effects. Our
numerical calculations confirm this picture. Since long range
interactions are important in establishing Wigner crystal-like
correlations,\cite{Schulz:prl93} screening from the lower wire acts to
inhibit somewhat Wigner crystal-like correlations. Longer range interactions
tend to make the peaks in the ground state density distribution and in
the density-density correlation more prominent.  Although the effect of
the decrease in physical density is small in the spinless case, we can
see in Fig.~\ref{fig:density_spinless} that distinct peaks in ground
state density are most prominent in the intermediate physical density
of eleven electrons per micron. Fig.~\ref{fig:density_spinless}
also shows that at a density of eleven electrons per micron, the
electrons more strongly avoid each other than at both higher and lower
density. Nevertheless, even at intermediate densities our numerical
results show that with experimental parameters of Steinberg \textit{et
al.},\cite{Steinberg:sci04,Steinberg:sci05} the interaction effect would be quite weak
for spin polarized electrons.

We note that numerical results for the electron density and
correlation function qualitatively similar to ours (inter-wire
screening effects not considered) were obtained in
Ref.~[\onlinecite{Szafran:prb04}].

\subsection{Momentum dependent tunneling matrix element}

Let us now consider tunneling between the interacting electron system
and a non-interacting electron system, shown schematically in
Fig.~\ref{fig:geometry}. Here we consider the situation when the
initial and final states of the interacting electron system (upper
wire) are both ground states. As discussed in Sec.~\ref{sec:model}, the tunneling
probability is then proportional to the matrix element $|M(k)|^2$.
\begin{figure}
\includegraphics[width=1.0\linewidth,clip=]{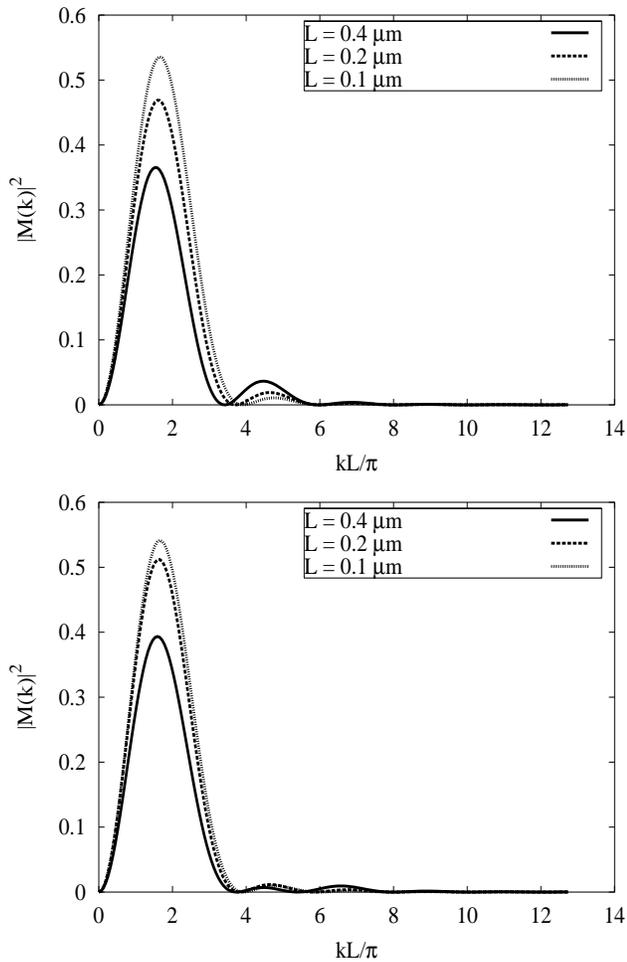}
\caption{\label{fig:three_spin} Top: Squared matrix elements of tunneling from a two to
    three electron system \emph{with} spin vs. scaled wavevectors $kL/\pi$.
	Here the $L=0.1\,\mu$m curve is indistinguishable from non-interacting limit. Bottom: Same as Top only tunneling from three to four electrons.
}
\end{figure}

\begin{figure}
\includegraphics[width=1.0\linewidth,clip=]{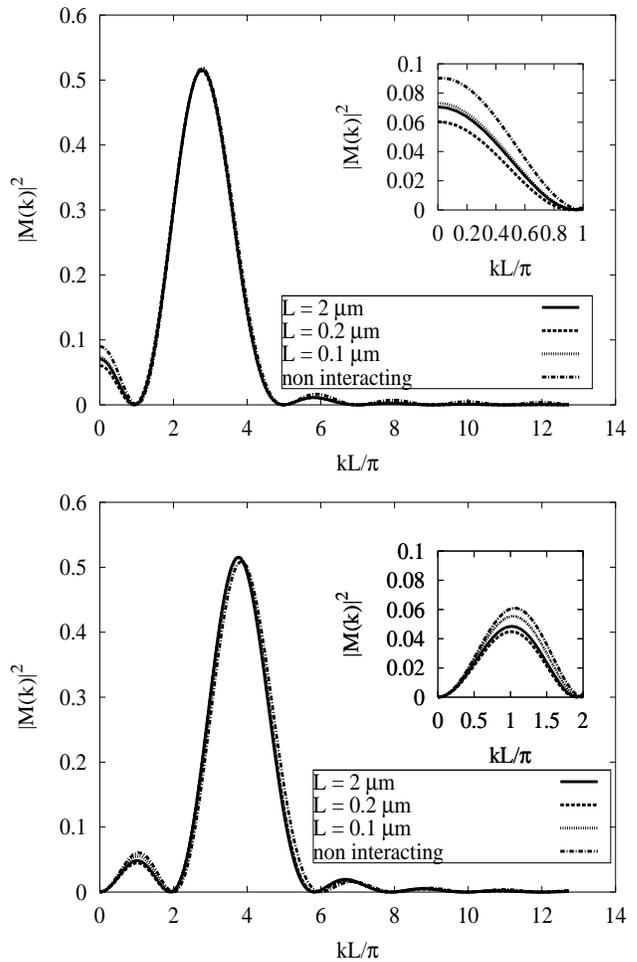}
\caption{\label{fig:three_spinless} Top:Squared matrix elements for tunneling from a two to
    three electron \emph{spinless} system vs. scaled wavevectors $kL/\pi$.
Bottom: Same as Top only tunneling from three to four electrons.}
\end{figure}

Typical results of the square of the absolute value of tunneling
matrix element $M$ are shown in Fig.~\ref{fig:three_spin} for
electrons with spin and in Fig.~\ref{fig:three_spinless} for electrons
without spin. All results are symmetric under the interchange $k\to
-k$. Note that the case with spin shows a much greater change with
density, although this appears to be limited mostly to the amplitude
of the peak in the tunneling.

It is clear from the discussion of previous passages that the spinless
system ground state under any density can be effectively described
using a non-interacting model. This is clearly seen in the plots. In
the spinless model, to get the noninteracting matrix element tunneling from $N-1$ to
$N$ electron ground state we put $k_{\tilde N}=\pi N/L$ in Eq~(\ref{eq:M1}). 
Manifestly, this has
the consequence that $M(k=0)$ is zero for even $N$ and non-zero for odd
$N$, and this is seen in the numerical result. There is also a sharp
maximum near $k_{\tilde N}=\pi N/L$. As $N\rightarrow \infty$, this maximum
will approach $k_{\tilde N}$. All these features are seen in
Figs.~\ref{fig:three_spinless}.  The
insets of the two plots further demonstrate what has already been
discussed at the end of the previous subsection: Electron interactions
cause greatest deviation of $|M(k)|^2$ from its non-interacting limit
at intermediate density. At both very large and small density
$|M(k)|^2$ approaches the non-interacting limit.

For systems with spin, the interaction has a more significant effect
on the matrix element.  Yet the salient feature here remains that,
even at very low density and strong interaction, and even after the system
changes from a Fermi liquid-like state to a Wigner crystal-like state,
the qualitative features of the scaled matrix element remain similar to
those of non-interacting solutions:  The positions of the maxima
are only slightly shifted, the positions of zeros a bit more so
(apart from $k=0$), and
the values of the maxima are reduced by no more than $30\%$ relative
to the non-interacting case, for densities down to $1\,\mu{\rm m}^{-1}$.
Furthermore, since for $N=3$ and $N=4$, the systems with spin have
the same matrix element in the non-interacting limit, the overall
features of the tunneling matrix elements, after scaling, are very
similar, even with interactions.  

\subsection{Spin exchange energy} 
\label{see}

\begin{figure}
\includegraphics[width=1.0\linewidth,clip=]{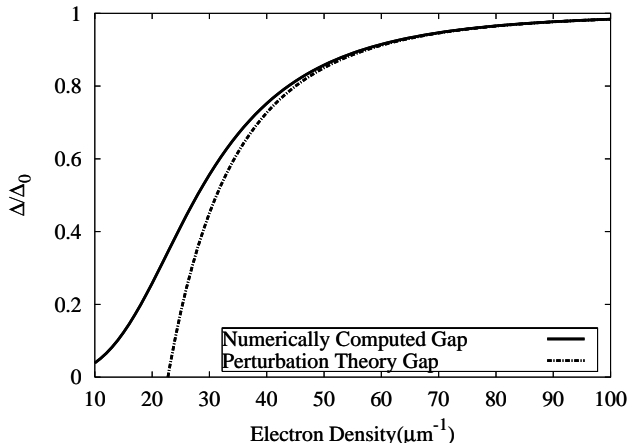}
\caption{\label{fig:gap_estimate} Comparison of the gap $\Delta$ 
    between the ground state and the first 
    excited state energy in a system of two electrons in a box, obtained by
	direct calculation versus by first order perturbation theory. 
	The unit  is $\Delta_{0}$, the non-interacting energy gap between the ground 
	and the first excited state. }
\end{figure}

In the model with spin, the first excited state has total spin $S=1$
for even numbers of electrons, and while for odd numbers of
electrons the situation can be more complicated.\cite{Hausler:zpb96} 
The energy gap between the ground state and the first
excited state decreases as the electron density tends to zero. The
high density, low interaction part of this change in the gap can be
understood by simple first order perturbation theory using a
non-interacting wavefunction, as shown in Fig.~\ref{fig:gap_estimate}.
In the low density, strongly interacting Wigner crystal-like
state, electrons are spatially confined around their equilibrium
positions. The system can be approximated as a Heisenberg spin chain,
where spin exchange is achieved by tunneling between adjacent
sites:\cite{Hausler:prb02,Matveev:prb04,Matveev:prl04}

\begin{figure}
\includegraphics[width=1.0\linewidth,clip=]{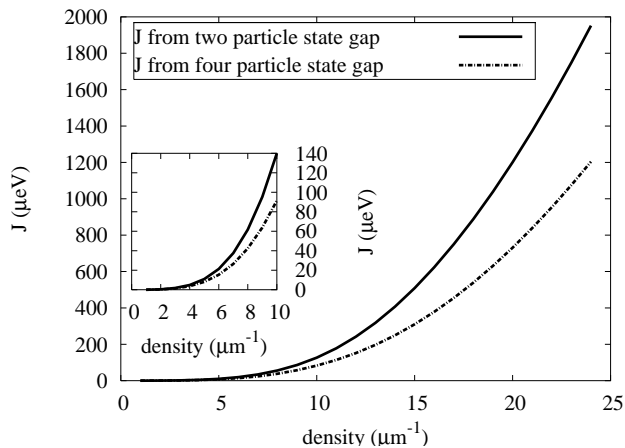}
\caption{\label{fig:gap_comp} Estimate of the spin exchange energy \emph{per
    electron} from a two and four electron system using the interaction in Eq.~(\ref{eq:int}).
    Both of the axes are in physical
    units: the x-axis is the physical density, in units of one electron per 
	$\mu$m, the y-axis is the physical exchange energy $J$ in units of $\mu$eV.  Typical experimental temperatures are in the range 250 mK-2 K $\approx 25  \mu $eV-200 $\mu$eV.
}
\end{figure}

\begin{equation}
\label{eq:H_spin}
H_{\rm spin}=J \sum_{i=1}^{N-1}\vec{S}_{i}\cdot\vec{S}_{i+1}\;. 
\end{equation}
This Hamiltonian describes the energy spectrum of low-lying spin
excitations.  Specifically, in the case of $N=2$ we simply have
\begin{equation}
{\label{equ:N2}} 
J=\Delta \;,
\end{equation}
where $\Delta$ is the gap between the ground state and the first excited 
state. For $N=4$ numerical solution of the four spin Heisenberg chain gives:
\begin{equation}
{\label{equ:N4}}
J=1.51777 \Delta\;.
\end{equation}
Analytical estimates of $J$ have been obtained in 
Refs.~[\onlinecite{Matveev:prl04,Matveev:prb04,Hausler:zpb96,Klironomos:cm05}].  
Hartree-Fock estimates are given in Ref.~[\onlinecite{Hausler:prb02}].
Related estimates of the spin velocity are given in 
Ref.~[\onlinecite{Kecke:prb04}] based on the ladder 
approximation and in Ref.~[\onlinecite{Creffield:epl01}].

Because the tunneling barrier between adjacent sites increases with
larger electron-electron interaction strength, the effective
Heisenberg exchange constant, and hence the gap between the ground
state and the first excited state, will decrease to zero as the
physical density goes to zero. It is interesting to note that the
effective Heisenberg exchange energy, extracted from the two-electron
system and four-electron system gap using Eqs.~({\ref{equ:N2}) and
({\ref{equ:N4}), closely follow each other in the low density regime,
as seen in Fig.~\ref{fig:gap_comp}. This supports our assumption that
spin excitations in the low-density regime are well described by a
Heisenberg model, and it suggests that in this regime, an estimate of
the Heisenberg exchange from an exact calculation of systems with a
few electrons can provide a reasonable prediction for the spin
velocity of an infinite wire with the same electron density. It 
should be noted, however, that the numerical value of $J$ may
depend sensitively on both short distance and long distance cut 
offs.\cite{Hausler:prb93}

\subsection{Mixed spin states}
\label{mss}

In the low density regime, where the spin exchange energy is small,
one may encounter an experimental regime where the temperature is
larger than $J$, but still small compared to the lowest energy for
charge excitations.  In that case, the tunneling conductance will be a
superposition of the results obtained from different spin states.
Also, under experimental conditions, the Zeeman energy $E_Z = g_s\mu_B
B$ may be comparable to or larger than $J$. (Here $\mu_B$ is the Bohr
magneton and $g_s$ is the $g$-factor of the electron spin.)  If the
Zeeman energy is large compared to both $k_BT$ and $J$, then the electron
system will be fully spin polarized, and calculations for spinless
electrons will apply.

As a simplest example, consider the case $N=2$, in the low-density regime.  If  the temperature and  the singlet-triplet splitting $\Delta$ are both small compared  to all charge-excitation energies in the $N=1$ as well  as in the  $N=2$ states,  then the sum  in Eq. (\ref{eq:BT}) may be  restricted  to six states:  the lowest singlet and triplet states of the $N=2$ system and the two spin states of $N=1$ ground state. As there are only two independent  matrix elements  involved, the sum may be further simplified  to read 
\begin{equation}
\mathcal{B}(k) = w_s |M_s(k)|^2  + w_t |M_t(k)|^2 ,
\end{equation}
where $M_s(k)$ and $M_t(k)$ are the matrix elements for the singlet and triplet  states defined in Eq.(\ref{eq:M}).  The weights $w_s$ and  $w_t$ depend  on the Zeeman energy $E_{\rm{Z}}$ and the chemical potential $\mu$, as  well as on $\Delta$ and $T$.  

In the limit $T \to 0$, the ratio $w_t/w_s$ will approach either zero
or infinity, depending on whether $E_{\rm{Z}}$ is smaller or larger
than $\Delta$. In the opposite limit, when $k_BT$ is large compared to
$E_{\rm{Z}}$ and to $\Delta$, we have $w_t/w_s \to 3/2$.  In
Fig. ~\ref{fig:FT12}, we show the momentum dependence of
$\mathcal{B}(k)$ for these three limiting values of $w_t/w_s$, at
$L=0.4\,\mu$m.

To illustrate further, we  may consider  the case where $\Delta$ is small compared  to $E_{\rm{Z}}$  and to $k_BT$, but the  ratio $E_{\rm{Z}}/(k_BT)$ is  finite.  If $\mu$ is adjusted so that the upper  wire has equal probability of having  $N=1$ or $N=2$ electrons,  then we  find
\begin{equation}
\frac{w_t}{w_s} = \frac{(2x^2+5x+2)}{6x},
\end {equation}
where $x=\exp[E_{\rm{Z}}/(k_BT)]$.  As expected,  we see that $w_t/w_s = 3/2$, for $x=1$,  and $w_t/w_s \to \infty$, when $x \to \infty$.

\begin{figure}[ht]
\includegraphics[width=1.0\linewidth,clip=]{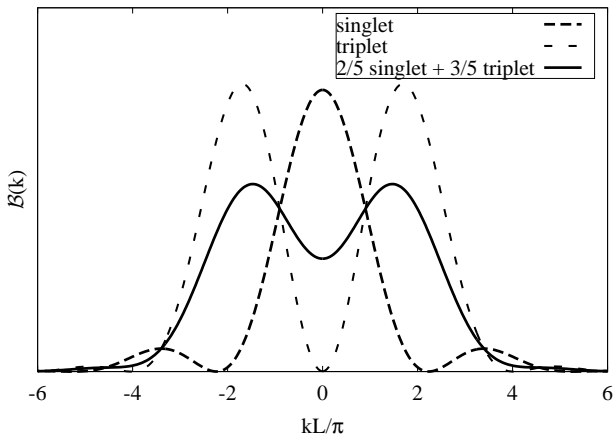}
\caption{\label{fig:FT12}
Momentum-dependence of tunneling conductance between one-electron and 
two-electron states for a wire of length $L$=0.4$\mu$m.  Dashed curve shows 
line-shape $|M_s(k)|^2$ for tunneling into the singlet ground 
state of the two particle system, applicable  at $T=0$ when the Zeeman energy 
$E_Z$ is smaller than the exchange gap $\Delta$. Short-dashed curve shows $|M_t(k)|^2$, for tunneling into the triplet ground state, applicable at $T=0$ 
when $E_Z > \Delta$.  Solid curve is a weighted average, applicable if $T $ is 
large compared to both $E_Z $ and $\Delta$ but small compared to the energy of the 
lowest charge excitation.}
\end{figure}

\section{Limit of large $N$}
\label{sec:largeN}

Properties of an infinite one-dimensional electron system have been
explored by a variety of techniques, including solutions of exactly
solvable models, renormalization group methods, bosonization, and
conformal field theory.  These techniques can be also adapted to the study of finite systems, when $N$ is large.  

In this section, we will use mostly
bosonization techniques to explore both spin-coherent and spin-incoherent
regimes.  Bosonization techniques give most readily the electron Green's functions
\begin{equation}
\mathcal{G}_\sigma(x,x^\prime,t)=\langle\psi_\sigma(x,t)\psi_\sigma^\dagger(x^\prime,0)\rangle
\label{Gd}
\end{equation}
as a function of position and time $t$.  For an infinite system, position variables  enter only in the combination $(x-x^\prime)$, but for a finite system, the distance from the ends may also be important.  To obtain the spectral density at a finite energy $\hbar \omega$ (setting $\mu=0$), one must take the Fourier transform with respect to $t$; however one may generally obtain an estimate of this by evaluating $G$ at an imaginary time $\tau$ equal to $\omega^{-1}$.   To obtain the ground-state tunneling amplitude for a finite system, when $k_BT$ is small compared to all excitation energies, we need to study the Green's function for $\tau \to \infty$. In particular, if the energies are adjusted so that $E^N=E^{N-1}$, then evaluating (\ref{Gd}) at a $(N-1)$-electron ground state yields
\begin{equation}
\lim_{\tau \to \infty} \mathcal{G}(x,x^\prime, \tau) = \Psi^N_{\rm{eff}} (x)   \Psi^{N\ast}_{\rm{eff}} (x^\prime) ,
\end{equation}
where  $\Psi^N_{\rm{eff}} (x)$ is the quasi-wavefunction defined in Eq.~(\ref{qwf}).  In the free-spin regime, at finite temperatures, we shall make use of a generalization of this result.

\subsection{Ground-state tunneling regime}

Numerous works have considered tunneling in the regime where the spin energy is
important.\cite{Kane:prl97,Fabrizio:prb95,Kleimann:prb02,Kleimann:prb00,yaro03,yaro02,Carpentier:prb02,Zulicke:prb02}
At sufficiently low energies, an infinite 
 interacting one-dimensional electron system is believed to be described as a
Luttinger liquid,\cite{Voit:rpp95} in which spin and charge
propagation are characterized by two distinct velocities,
$v_s$ and $v_c$. At finite bias voltages, the
amplitude for momentum resolved tunneling into a Luttinger liquid is
expected to show structure reflecting both of these velocities.
This follows from the form of the one-particle Green's function ($z=x-x^\prime$)
\begin{equation}
\mathcal{G}(z,\tau)\sim\frac{e^{ik_Fz}}{\sqrt{(v_s\tau-iz)(v_c\tau-iz)[(v_c\tau)^2+z^2]^\alpha}}+{\rm c.c.}\,,
\end{equation}
dropping a short-distance cutoff-dependent prefactor.
In the limit where the bias voltage $V \to 0$ and $T \to 0$, however, the
momentum-resolved tunneling conductance reduces to $\delta$ functions at $k = \pm k_F$
similar to that of a non-interacting electron system.  The primary
effect of the interactions is then to reduce the amplitude of the
$\delta$-functions. For an infinite system, the amplitude is predicted to vanish as $\max(k_BT,V)^{\alpha}$, when $T$ and $V$ approach 0, where the bulk tunneling exponent $\alpha \ge 0$
depends on the strength of the electron-electron interaction. Tunneling into a point near the end of a semi-infinite system is characterized by a different exponent $\alpha_{\rm{end}}$, which is generally larger than $\alpha$. The tunneling exponents are related to the Luttinger liquid interaction parameter $g$
by $\alpha=(g + g^{-1} -2)/4$ and $\alpha_{\rm{end}}=(g^{-1} - 1)/2$ for a single mode wire.
\cite{Voit:rpp95} For fermions with repulsive interactions, as considered here, one has $g<1$. For translationally-invariant systems, $g=v_F/v_c$, where $v_F$ is the Fermi velocity with turned off interactions. For not too strong interactions, one can make an RPA estimate for $g$ of a cylindrical wire of radius $R$ with electron density $\bar n$ at a distance $D$ from a two-dimensional screening gate:\cite{Kane:prl97}
\begin{equation}
g^{-1}\approx\sqrt{1+\frac{8\ln(2D/R)}{\pi^2\bar na_B}}\,.
\end{equation}
We remark that if the lower high-density wire had charge modes propagating with a very different velocity from the upper wire, the interwire interactions would not significantly couple or modify the charge modes of individual wires.\cite{Matveev:prl93} Using parameters of Ref.~[\onlinecite{Steinberg:sci04,Steinberg:sci05}], $2D\approx1\,\mu$m, $R\approx20$~nm, $a_B\approx10$~nm, and $\bar n\approx30\,\mu$m$^{-1}$ near the localization transition, we get $g\approx0.3$ which is comparable to the measured $g\approx0.5$.

Kane, Balents, and Fisher\cite{Kane:prl97} considered tunneling into a finite-length metallic carbon nanotube, which has four excitation modes, rather than the single charge and spin modes considered here.  Nevertheless, their formulas may be readily adapted to the present case.  The general result, for a wire with uniform electron density and hard-wall confinement at the ends,  may be written in the form 
\begin{equation}
\label{Psialph}
\Psi^N_{\rm{eff}} (x) \sim \frac{1}{\sqrt{LN^\alpha}} 
 \left[\sin\left(\frac{\pi x}{L}\right)\right]^{\frac{1}{2}(\alpha_{\rm{end}}-\alpha)} \sin(k_F x)\;,
\end{equation}
valid for positions not too close to the wall where the factors involving
$\alpha$ cease to be a good approximation. 

As $\alpha$  increases with increasing  interaction strength, the overall amplitude of the 
tunneling matrix element $M(k)$, obtained by Fourier transforming (\ref{Psialph}),  will 
decrease rapidly with increasing interaction strength, for $N$ large but fixed.  However, the momentum-dependence remains qualitatively similar to  the form (\ref{eq:M1}) for non-interacting electrons, and the peak at $k_F$ broadens only gradually with increasing interaction strength, as illustrated in  
Fig.~\ref{fig:large_N}.

\begin{figure}[ht]
\includegraphics[width=1.0\linewidth,clip=]{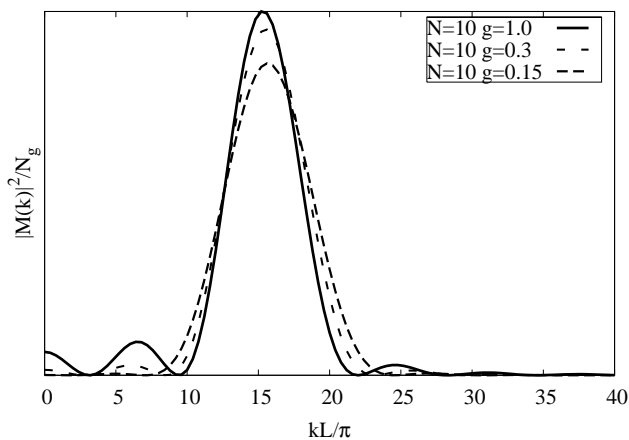}
\caption{\label{fig:large_N} Momentum-dependence of ground-state tunneling for N=10 electrons and various interaction strengths, obtained by calculating  $|M(k)|^2$ from the effective wavefunction (\ref{Psialph}). Each curve has been divided by a normalization constant $N_g$, chosen so that the plotted curves have equal areas. The norm $N_g$ decreases rapidly with  increasing interaction strength (decreasing $g$), but the width of the peak broadens only slightly. Note
that for $g=1$ the results are equivalent to the $\tilde N = 5$ case in
Fig.~\ref{fig:FT01_box}.}
\end{figure}

Tserkovnyak {\em et al.}\cite{yaro02,yaro03} have explored the
implications of Luttinger liquid theory for momentum-resolved tunneling into a long
finite wire with {\em{soft}} confinement at the ends.  Although the emphasis
of that work was on tunneling at finite bias, where the discreteness
of the energy levels plays a relatively minor role, the analysis can also be applied to estimate the form  of $\Psi^N_{\rm{eff}} (x)$ and the 
matrix element $M(k)$ for 
the ground-state tunneling into a wire with large $N$, for not too strong interactions. Essentially, the factor $\sin (k_F x)$ in (\ref{Psialph}) should be replaced by the WKB wavefunction for a non-interacting electron in a self-consistent potential that gives the correct density profile along the wire. The zeroes of $ \Psi^N_{\rm{eff}} (x)$ will be spaced further apart near the ends of the wire than near the center, but near the center of the wire, the wavefunctions will look quite similar for the cases of soft and hard confinement.  For increasing values of the  the electron-electron interaction, the envelope factor in (\ref{Psialph}) is increasingly weighted towards the center of the wire, so that the Fourier transform $M(k)$ should become correspondingly  insensitive to the difference between soft and hard confinement.

\subsection{Finite temperatures}
\label{sec:finite_T}

At finite temperatures $T$, for the infinite system, one expects that
the $\delta$-functions at $k=k_F$ will be broadened by an amount of
order $\delta k \approx k_B T / (\hbar v_s)$.  If the temperature is
larger than the energy of the lowest spin wave mode, $\pi \hbar v_s /
L $, the broadening will be larger than $2\pi/L$ for a finite system
of length $L$.  The width $\delta k$ becomes comparable to $k_F$ when
$k_B T$ becomes comparable to $J\sim\hbar v_s k_F$, 
the energy of the shortest wavelength spin excitations.

If the spin velocity is very much smaller than the charge velocity,
one can enter a regime where $k_B T$ is larger than the energy of the
shortest-wavelength spin excitations, but still low enough that
relatively few short-wavelength charge modes are excited. This regime has been explored
in recent papers by Cheianov and Zvonarev,\cite{cheianov03,Cheianov04}
and by Fiete and Balents.\cite{Fiete:prl04} The form factor for
momentum-resolved tunneling is indeed broadened by an amount of order
$k_F$ in this regime, as the Green's function in position
space is found to fall off exponentially with distance. A more precise
description of this behavior will be given in the next subsection.

The lowest energy cost for a charge excitation is $\epsilon_c = \pi
\hbar v_c/L$.  The highest energy for a spin state with multiple
excitations is $ \approx N \hbar v_s k_F $.  Earlier, we defined an
extreme free-spin regime as having a temperature high enough so that
all spin states have equal weight, while there are no charge
excitations. This makes sense for a small system, but is very
restrictive for a large system, as it requires that $v_s/v_c \ll 1/N^2
$.  However, the spins will actually be effectively free as long as
$k_B T$ is greater than the energy of a single short wavelength
spin-excitation, while smaller than the lowest charge excitation.
This only requires that $v_s/v_c \ll 1/N $.  The results for the
momentum dependence of the tunneling will be the same in this case as
for the extreme free-spin regime.

\subsection{Free-spin regime}
\label{sec:effective_theories}
 
We first focus on the regime with energy-scale hierarchy $J\ll
k_BT \ll \hbar v_c k_F $,  so that all spin configurations have
effectively equal statistical weight while \textit{short-wavelength}
charge excitations are absent in equilibrium. (We do not yet impose the more restrictive condition   $k_B T \ll \epsilon_c$.) We are particularly
interested in the Green's function ${\cal G}(x,x',\tau)$ at large
imaginary times $\tau$ but spatial
 separations of the order of the inter-electron spacing, as these are
 the distances that dominate the tunneling response,
as discussed in the following.  For
 times smaller than the time for a charge excitation to cross the
 system, ${\cal G}$ should depend only on the separation $x-x'$. For
 longer times, there may also be a dependence on the distance from the
 boundary.  We will first calculate
 the Green's function in the infinite wire limit.

A quantity that will play a crucial role in our analysis of the tunneling at $\tau\gg a/v_c$ is the root-mean-square displacement of an electron during a time $\tau$ due to quantum fluctuations [derived in Eq.~(\ref{eq:G_0})]:
\begin{equation}
\bar u(\tau)=\frac{a}{\pi}\sqrt{2 g \ln(v_c\tau/a)}\;,
\label{eq:u_bar}
\end{equation}
where $a$ is the mean spacing of the electrons. When $\bar u(\tau)\ll a$ (which is possible only for very strong interactions, $g\ll1$), electrons can be thought of as forming a ``true Wigner crystal'' on the relevant energy scales. When $\bar u(\tau)\gtrsim a$ (which for strong interactions corresponds to exponentially small energies), fluctuations destroy the long-range translational order, but significant Wigner crystal correlations may be present at intermediate length scales.  We refer to this situation as the ``fluctuating Wigner crystal''. For a finite size system, the fluctuations $\bar u(\tau)$ are cut-off by the length of the system at times $\tau=L/v_c$. The momentum structure 
of the tunneling
will exhibit qualitative differences depending on the relative size of $a$ and
$\bar u$, and the form of the Green's function itself will also depend on the 
relative sizes of $a$,  $\bar u$, and $x-x'$.

The condition $J\ll k_BT\ll \hbar v_c k_F $ requires that the short-distance repulsion between electrons be very strong\cite{cheianov03,Matveev:prl04}, so that neighboring electrons exchange positions only on a time scale $h/J$ which is longer than $\tau\sim h/(k_B T)$. The Green's function ${\cal G}$ can then be expressed as a series of the form:
\begin{equation}
{\cal G}(x,x',\tau)= \sum_{jm} {\cal G}_{jm} (x,x',\tau) (-1)^m  p_m\;,
\label{eq:G_Bert}
\end{equation}
where ${\cal G}_{jm}$ is the Green's function for  spinless bosons which
do not change places, when a particle is inserted at time 0 at space point $x'$ as the 
$j^{\rm th}$ particle on the lattice,  
and removed at time $\tau$ at point $x$ in lattice position $j+m$.
The factor $(-1)^m$ encodes
the Fermi statistics associated with the permutation of a given electron through
$m$ intermediate electrons, and if we were to set $p_m=1$, we would recover the Green's function for spinless fermions.  Here, however we set  
$p_m=2^{-|m|}$, which is the probability of finding $|m|+1$ consecutive spins 
pointing in the same direction.  In the limit we are considering, where all spin states are effectively degenerate, and the spin dynamics
is effectively frozen out (exchange events do not occur), 
  all electrons between lattice positions $j$ and $j+m$ must have parallel spins to contribute to the Green's function 
  ({\it{i.e.,}} for the world line trajectories to wrap around the imaginary time torus,
as discussed in Ref.~[\onlinecite{Fiete:prl04}]).

Let us first consider the situation where $|x-x'|$ is much larger than
$a$ and $\bar u$. 
The creation  operator for a spinless fermion  can be bosonized in terms of
$\theta$ and $\phi$ fields associated with the fluctuations in the 
electron density and momentum density.

Specifically,  $\theta$ is related
  to the particle density fluctuations through the familar
  relationship $\delta n(x)=\frac{1}{\pi}\partial_x\theta(x)$, and the fields 
  satisfy $[\phi(x),\theta(x')]=-i\pi H(x-x')$ where $H$ is the Heaviside 
  function.  
In a first-quantized path integral formulation of the Green's function, 
the dominant 
contributions come from those world-line trajectories that
fluctuate only a small amount from configurations corresponding to the mean
density (due to strong Coulomb repulsion in the Wigner crystal limit), 
as discussed in
Ref.~[\onlinecite{Fiete:prl04}].

For $a,\bar u\ll|x-x'|\ll L$ and $x,x'$ away from the wire boundary, 
density fluctations can be 
treated as a continous variable and the Green's function   (\ref{eq:G_Bert})  
may be expressed as\cite{Fiete:prl04}
\begin{equation}
\label{eq:G_start}
{\cal G}(x,x',\tau) \sim \langle 2^{-|N(x,x',\tau)|} (-1)^{N(x,x',\tau)} e^{i[\phi(x,\tau)-\phi(x',0)]}\rangle\;.
\end{equation}
Here $N(x,x^\prime,\tau)$ is the particle number operator between points $x$ and $x^\prime$. 
The first factor, in (\ref{eq:G_start}) corresponds to $p_m$, while the second and third together represent the product of creation and annihilation operators for a spinless fermion.  
The average is taken at $T \rightarrow 0$ in the charge
sector.
Using the bosonized expression
of the particle number in a distance $x-x'>0$,
\begin{equation}
N(x,x',\tau)=\bar n (x-x') + \frac{1}{\pi}\left[\theta(x,\tau)-\theta(x',0)\right]\;,
\label{eq:N}
\end{equation}
the electron number can be written in terms of the
average density, $\bar n$, and a 
fluctuating piece expressed in terms of the $\theta$ fields. 
Using Eq.~(\ref{eq:N}), and approximating operator $(-1)^{N}$ as ${\rm Re} [ e^{i\pi N}]$, the Green's function can be expressed as ${\cal G}={\cal G}_+ + {\cal G}_-$, where\cite{Fiete:prl04}
\begin{align}
{\cal G}_+(x,x',\tau) \sim&  e^{-\tilde{k}_F (x-x')\frac{\ln 2}{\pi}}
e^{i\tilde{k}_F(x-x')}  
\langle e^{-\frac{\ln 2}{\pi}[\theta(x,\tau)-\theta(x',0)]}
\nonumber \\
&\times e^{i[\theta(x,\tau)-\theta(x',0)]} e^{i[\phi(x,\tau)-\phi(x',0)]}\rangle\;,\label{eq:G_gen}
\end{align}
and $\mathcal{G}_-$ (as we will see from the derivation below) is given by ${\cal G}_- = [{\cal G}_+]^*$. The first two factors in Eq.~(\ref{eq:G_gen}) come from the exponentiation of the average
density and we have defined $\tilde{k}_F\equiv\pi \bar n$.  The exponential 
decay of the first factor results from spin
averaging and the oscillatory second term comes from Fermi
statistics, as discussed above. 

We now compute the part of the
Green's function coming from fluctuations in the charge sector 
using the effective low energy ``elastic'' action
\begin{equation}
S_c=\int dx d\tau \frac{v_c}{2\pi}\left[\frac{1}{2g}(\partial_x \theta)^2+2g(\partial_x \phi)^2 \right]+\frac{i}{\pi}\partial_\tau\phi\partial_x\theta,
\label{eq:action}
\end{equation}
obtained from the Hamiltonian
\begin{equation}
{\cal H}=\int dx \frac{v_c}{2\pi}\left[\frac{1}{2g}(\partial_x \theta)^2+2g(\partial_x \phi)^2 \right]\;,
\label{eq:H_bos}
\end{equation}
A justification of the Harmonic form (\ref{eq:H_bos})
for a strongly interacting system has been given in 
Refs.~[\onlinecite{Matveev:prb04}]  and ~[\onlinecite{Haldane:prl81}].  
Since the action is quadratic, the average can be moved to the exponent 
\begin{align}
{\cal G}_+(x,x',\tau) 
\sim&e^{-\tilde{k}_F(x-x') \frac{\ln 2}{\pi}}e^{i\tilde{k}_F(x-x')} e^{-\frac{1}{2}\left( 1+i\frac{\ln 2}{\pi}\right)^2 \langle \Theta^2\rangle}\nonumber \\
&\times e^{-\frac{1}{2}\langle \Phi^2\rangle } e^{-\left(1+i\frac{\ln 2}{\pi}\right)\langle \Phi \Theta \rangle }\;,
\label{eq:G_bos}
\end{align}
where $\Phi=\phi(x,\tau)-\phi(x',0)$ and $\Theta=\theta(x,\tau)-\theta(x',0)$.
If we assume as before that the electrons are confined to a wire of length
$L$, the fields must satisfy the boundary conditions 
$\partial_x\phi(x=0,L)=0$ and $\theta(L)-\theta(0)=\pi (N-\bar n L)$.
Here $N$ is the total number of electrons.  Expanding the fields in a 
Fourier series gives\cite{Kane:prl97}
\begin{eqnarray}
\label{eq:theta_FT}
\theta(x,\tau)=\sum_{m=1}^\infty i \sqrt{\frac{2g}{m}}\sin\left(\frac{m\pi x}{L}\right)(b_m e^{-\omega_m \tau}-b^\dagger_m e^{\omega_m \tau})\nonumber \\+\theta^{(0)}(x)\;\;\;\;
\end{eqnarray}
\begin{eqnarray}
\label{eq:phi_FT}
\phi(x,\tau)=\sum_{m=1}^\infty\sqrt{\frac{1}{2gm}}\cos\left(\frac{m\pi x}{L}\right)(b_m e^{-\omega_m \tau}+b^\dagger_m e^{\omega_m \tau})\nonumber \\+\Phi_c\;,\;\;\;\;\;
\end{eqnarray}
where the zero mode term $\theta^{(0)}(x)\equiv \frac{x}{L}\pi (N -\bar n L)$, the $b_m$ satisfy $[b_m,b^\dagger_{m'}]=\delta_{mm'}$ and the operators
$N$ and $\Phi_c$ satisfy $[N,\Phi_c]=1$.  These field expansions
diagonalize the Hamiltonian (\ref{eq:H_bos}) of the charge sector as
\begin{equation}
H=\frac{E_C}{2}N^2 +\sum_{m=1}^\infty m \epsilon_c b^\dagger_m b_m\;.
\end{equation}
The first term is the ground-state energy in the charge sector, with the charging energy 
$E_C = \epsilon_c / (2 g)$. 
The second term represents the energy associated with collective density
oscillations of the low density electron gas.

Continuing with the evaluation of the Green's function for $z=x-x' \gg a, \bar u$, 
it is useful to first consider the limit of $L\to \infty$. 
Then the correlators in the exponent of (\ref{eq:G_bos}) 
depend only on $z$ and can be evaluated as
$\langle
\Theta^2\rangle=g \ln[(z^2+v_c^2\tau^2)/a^2]$, $\langle \Phi^2\rangle =
(4g)^{-1}\ln[(z^2+v_c^2\tau^2)/a^2]$ and $\langle \Phi \Theta \rangle =
2^{-1}\ln[(v_c\tau -iz)/(v_c\tau+iz)]$. 
(Here $a = \bar n^{-1}$ is the average spacing between electrons.)
The resulting Green's function $\mathcal{G}=\mathcal{G}_++\mathcal{G}_-$ was computed
earlier:\cite{cheianov03,Fiete:prl04}
\begin{equation}
{\cal G}(z,\tau)=\frac{C'e^{-\tilde{k}_F |z| \frac{\ln 2}{\pi}}}{(z^2+v_c^2\tau^2)^{\Delta_g}}\left[\frac{e^{i(\tilde{k}_F z -\varphi_g^+)}}{v_c \tau-iz}+\frac{e^{-i(\tilde{k}_F z -\varphi_g^-)}}{v_c \tau+iz}\right]
\label{eq:G_asym_x}
\end{equation}
where $C'$ is an undetermined constant.
The phases $\varphi_g^\pm$ are given by
\begin{equation}
\varphi^\pm_g(z,\tau)=\frac{\ln 2}{\pi}\left [ g \ln\left(\frac{z^2+v_c^2\tau^2}{a^2}\right)\pm \frac{1}{2}
\ln \left(\frac{v_c\tau-iz}{v_c\tau+iz}\right) \right]\;.
\label{eq:phases}
\end{equation}
The power law decay of (\ref{eq:G_asym_x})
is characterized by the {\it anomalous} exponent
\begin{equation}
\Delta_g=\frac{1}{8g}+\frac{g}{2}\left[1-\left(\frac{\ln 2}{\pi}\right)^2\right] -\frac{1}{2}\;.
\label{eq:D_g}
\end{equation}
We note that the Green's function (\ref{eq:G_asym_x}) is valid only for
$z\gg a,\bar u$ and that it decays very rapidly with distance due to the 
incoherent spin degrees of freedom.  The exponential piece in 
(\ref{eq:G_asym_x}) shows that the spin incoherent regime ($J\ll k_BT$)
destroys the strictly power law decay of the correlation function present
in the Luttinger liquid theory where the spin degrees of freedom are
coherent. 
In order to calculate the momentum resolved tunneling, we shall need an alternate formula for 
${\cal G}(z,\tau)$, which reduces to (\ref{eq:G_asym_x})  when $a,\bar u \ll z \ll v_c \tau$, but is also valid when
$z \sim a,\bar u$.

When the distance $z$ is comparable to the separation between electrons 
(where the density fluctuations are large compared to the average density), 
but $z \ll v_c\tau$ we have in the limit $L\to\infty$\cite{Fiete:prl04}
\begin{align}
{\cal G}(z,\tau)&\sim \sum_{j=-\infty}^\infty 2^{-|j|}(-1)^j \langle \delta(N(z,\tau)-j)e^{i[\phi(z,\tau)-\phi(0,0)]}\rangle \nonumber \\
&= \!\!\!\! \sum_{j=-\infty}^\infty \!\!\!\! 2^{-|j|}(-1)^j\int \frac{d\lambda}{2\pi} e^{-i\lambda (j-\bar n z)}
\langle e^{i[\lambda \Theta(z,\tau)/\pi + \Phi(z,\tau)]}\rangle \nonumber \\
&= \!\!\!\! \sum_{j=-\infty}^\infty \!\!\!\!\! 2^{-|j|}(-1)^j\!\!\! \int \!\!\! \frac{d \lambda}{2\pi} e^{-i\lambda (j - \bar n z)}
e^{-\frac{1}{2}\left(\frac{\lambda^2}{\pi^2} \langle \Theta^2\rangle 
+2\frac{\lambda}{\pi}\langle \Theta \Phi \rangle +\langle \Phi^2 \rangle \right) } \nonumber \\
&\approx \sqrt{\frac{\pi}{2 \langle\Theta^2\rangle}}\sum_{j=-\infty}^\infty  2^{-|j|}(-1)^je^{\frac{-\pi^2(\bar n z-j)^2}{2 \langle\Theta^2\rangle}} e^{-\langle \Phi^2\rangle/2}
\nonumber \\
&= \frac{a}{\sqrt{2\pi}\bar u(\tau)} \left(\frac{a}{v_c\tau}\right)^{\frac{1}{4g}} f\left(z, \bar u(\tau)\right) \;,
\label{eq:G_0}
\end{align}
where
\begin{equation}
\label{fdef}
f(z,\bar u) \equiv \sum_{j=-\infty}^\infty 2^{-|j|}(-1)^j  e^{-\frac{(z-ja)^2}{2 \bar u^2}}\;.
\end{equation}
In the second to the last line of (\ref{eq:G_0}), we have made use of the condition 
$z \ll v_c\tau$ to neglect the contribution of the correlator
$\langle \Theta \Phi \rangle$, and
in the last line we have evaluated the
remaining correlators and used the definition (\ref{eq:u_bar}) for $\bar u$.

With Eqs.~(\ref{eq:G_asym_x}) and (\ref{eq:G_0}) in hand, we are now ready
to study the momentum-resolved tunneling spectrum of a long quantum
wire in the limit $J\ll k_B T\ll \hbar v_c k_F$, by computing the Fourier transform 
${\cal G}(k,\tau) =\int dz e^{-ik z}{\cal G}(z,\tau)$.
We are interested in large values of $\tau$, so  the Fourier transform will be dominated by contributions from  $z \ll v_c \tau$. The dominant, short-distance, $z\sim a$, correlations that determine the momentum composition of the Green's function should thus be correctly described by Eq.~(\ref{eq:G_0}).
The Fourier transform of Eq.~(\ref{fdef}) can be computed directly,
but it is convenient to express $f$ as the convolution of a Gaussian with the function 
\begin{equation}
l(z)=\sum_{j=-\infty}^\infty 2^{-|j|}(-1)^j \delta(z-ja)\;.
\end{equation}
The Fourier transform of $f$ is thus the product of a Gaussian and the Fourier transform of $l(z)$, which leads to
\begin{equation}
\label{GktA}
{\cal G}(k,\tau) \sim a \left(\frac{a}{v_c\tau}\right)^{\frac{1}{4g}} A\left(k, \bar u (\tau) \right)\;,
\end{equation}
\begin{equation}
\label{Adef}
A(k, \bar u ) \equiv e^{-\frac{k^2\bar u^2}{2}} \left [\frac{3}{5+4 \cos(ka)}\right ]\;.
\end{equation}.

The amplitude for momentum-resolved tunneling into an infinite wire, at an energy $\hbar \omega$, is  determined by the inverse Laplace transform of the  function  ${\cal G}(k,\tau)$.  In practice, this is determined by the form of ${\cal G}(k,\tau) $ for $ \tau \approx \omega^{-1}$.   For tunneling into the charge ground state, we shall actually want to take the limit $\tau \to \infty$, with the system length $L$ held finite.  We shall see below, that the behavior of  $\cal G$ in this limit can be obtained,  to a good approximation, by using the expressions for an infinite system, with $\tau $ replaced by $L/v_c$.  Thus the momentum dependence will be given by the function $A(k,\bar u)$ defined in (\ref{Adef}), with $\bar u$ evaluated as
\begin{equation}
\label{uL}
\bar u(L/v_c)=\frac{a}{\pi}\sqrt{2 g \ln(L/a)}\;.
\end{equation}

Equations (\ref{GktA}) and (\ref{Adef}) are,  therefore, central results of this section and
they have several features worth emphasizing.  The first is the
momentum structure: There is an exponential envelope centered about
zero momentum, $e^{-\frac{k^2\bar u^2}{2}}$, whose width is given by
the parameter $\bar u$ measuring the fluctuations of an electron's
position.  Larger fluctuations imply a more sharply peaked envelope in
momentum space.  This envelope multiplies another momentum dependent
function, which is sensitive to the mean spacing of the electrons and has maxima  
at $k=\pm \pi/a= \pm {\tilde{k}}_F$.   

Clearly, the overall shape of the momentum distribution will depend on
the relative size of $\bar u$ and $a$. The maxima at finite $k$ will
disappear and merge into shoulders for $\bar u / a$ larger than a
critical value of order 0.75. By contrast, in the limit $\bar u / a
\to 0$, the height of the maxima is nine times the value at $k=0$. The
momentum-dependence of the spectral function $A(k, \bar u)$ is shown
in Fig.~\ref{fig:smallx_FT_u} for several choices of $\bar u$.

\begin{figure}[ht]
\includegraphics[width=1.0\linewidth,clip=]{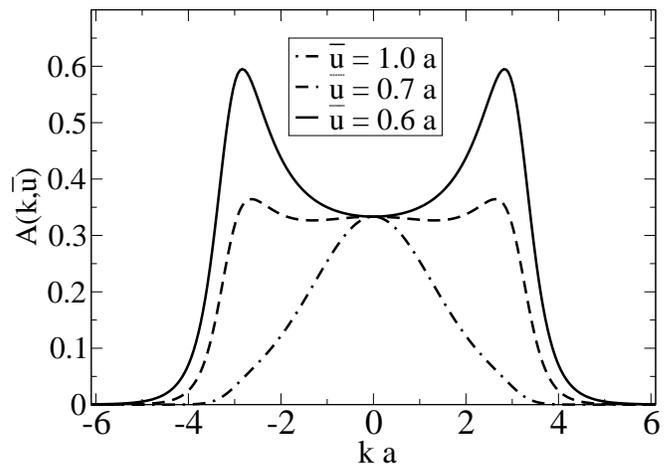}
\caption{\label{fig:smallx_FT_u} Spectral function $A(k, \bar u)$, which determines the momentum dependence of tunneling in the 
{\em{free-spin regime}}.  The quantity $\bar u$ is the root-mean-square electron displacement, due to quantum fluctuations, from the sites of a classical Wigner crystal, and $a$ is the lattice spacing. When $\bar u \gtrsim a$,
the momentum distribution is single lobed and peaked about zero momentum.
In the opposite limit, when $\bar u \ll a$, the momentum distribution exhibits a
doubled lobed structure with peaks near $k=\pm \tilde k_F = \pm \pi/a$.
The peaks have width slightly less than $\tilde k_F$, in agreement with Eq.~(\ref{eq:A_k}) obtained from the Green's function (\ref{eq:G_asym_x}).}
\end{figure}

One can also obtain the double-lobed structure of ${\cal G}(k,\tau)$ in the limit where $\bar u\ll a$ by Fourier transforming Eq.~(\ref{eq:G_asym_x}), although it is, strictly speaking, only a valid approximation when $z\gg a$, while it is the region $z\sim a$ which dominates the spectral properties. Neglecting the slowly-varying phases $\varphi_g^\pm$, Eq.~(\ref{eq:G_asym_x}) gives 
${\cal G}(k,\tau) \approx \mathcal{A}_++\mathcal{A}_-$, with
\begin{equation} 
\mathcal{A}_\pm(k) \propto \frac{1}{(k\pm\tilde k_F)^2+(\tilde k_F\frac{\ln 2}{\pi})^2}\;. 
\label{eq:A_k}
\end{equation} 

The density of states for tunneling into  a single point $x$, far from the ends of the wire, at a finite energy $\omega$, is approximately given by 
$\mathcal{A}(\omega) \approx  \tau \mathcal{G}(z=0, \tau)$, with $\tau = \omega^{-1}$.
Using (\ref{eq:G_0}), we see that when $\bar u $ is of order $a$ or larger, 
the frequency-dependence of the tunneling density of states is given by
$\mathcal{A}(\omega) \propto \omega^{\tilde{\alpha}}/ \bar u$, with\cite{Fiete:prl04} 
\begin{equation}
\label{alfs}
{\tilde{\alpha}}  = \frac{1}{4g} - 1 .
\end{equation}
This implies that for $4g>1$
the density of states {\em diverges} as $\omega$ decreases, {\em
unlike} the behavior of a Luttinger Liquid, while for $4g<1$ the
tunneling density of states is suppressed, qualitatively similar to
the behavior of a Luttinger Liquid.\cite{exponent_comment}

The density of states must be modified if one tunnels into a point near to the end of a  semi-infinite wire, so that the distance from the end is small compared to $v_c / \omega$. 
For $\tau > x/v_c$, the expectation values $\langle \Theta^2 \rangle$, $\langle \Phi^2 \rangle$,  and $\langle \Phi \Theta \rangle$, which enter (\ref{eq:G_0}) are affected by the proximity of the boundary. In particular, fluctuations in $\Theta$ are reduced, but  $\langle \Phi^2 \rangle$ is doubled near the end of the wire, leading to a reduction of the tunneling amplitude. One finds, for fixed value of $x$ and $\omega \to 0$, that the tunneling density of states scales as 
$\mathcal{A}(x, \omega) \propto \omega^{{\tilde{\alpha}}_{\rm{end}}} 
x ^{{\tilde{\alpha}}_{\rm{end}}-{\tilde{\alpha}}}$ , where
\begin{equation}
\label{alfsend}
{\tilde{\alpha}}_{\rm{end}} = \frac{1}{2g} - 1.
\end{equation}

We now discuss in greater detail the effects of finite system length $L$.  To obtain the resonant tunneling amplitude in the free-spin regime, we need to calculate the function 
\begin{equation}
\mathcal{G}(x,x^{\prime}) =  [ \, \sum_{\alpha \sigma} \langle\beta | \psi_\sigma (x) | \alpha\rangle
\langle\alpha| \psi_{\sigma}^{\dagger} (x^{\prime} )| \beta\rangle  \, ]_{\rm{ave} } .
\end{equation}
Here, $\alpha$ and $\beta$ denote the spin states of the system with $N$ and $N-1$ electrons, respectively, and the average is taken over the spin states $\beta$.  The tunneling conductance (\ref{eq:G}) is given by [see Eq.~(\ref{eq:CT})] 
\begin{equation}
\label{BF}
G_\pm \sim  \lambda^2 \beta\nu \int \,  dx \, dx^{\prime} e^{-ik_\pm(x - x^\prime) } \mathcal{G}(x,x^{\prime} )\; . 
\end{equation}
The function $\mathcal{G}(x,x^{\prime} )$ is determined by the function 
$ \mathcal{G}(x,x^{\prime}, \tau)$, in the limit $\tau \to \infty$, which we can evaluate using the same steps as in Eq.~(\ref{eq:G_0}) above. 
Because of the exponential dependence of the first factor in Eq.~(\ref{eq:G_bos}), we are only concerned with  situations with  $|x-x'| \ll L$. 
We must now take into account the discreteness of the normal modes, as well as the proximity of the boundary, with the result that the expectation values $\langle \Theta^2 \rangle$ and $\langle \Phi^2 \rangle$,   entering (\ref{eq:G_0}),
depend on $x/L$, as well as, logarithmically, on $L/a$.
The general result is
\begin{equation}
\label{Gfin}
\mathcal{G}(x, x^{\prime})  \sim  \frac{L^{-1} (a/L)^{\tilde{\alpha}}}{\sqrt{g \ln (L/a)}} 
\left[ \sin \left(\frac{\pi x}{L}\right) \right]^{{\tilde{\alpha}}_{\rm{end}} -  {\tilde{\alpha}}  } 
f(x-x^{\prime}) ,
\end{equation}
 where ${\tilde{\alpha}}$ and $  {\tilde{\alpha}}_{\rm{end}}$  are given by (\ref{alfs}) and (\ref{alfsend}), and
$f(x-x')$ is given by (\ref{fdef}) with $\bar u = \bar u(L/v_c)$ . 
(The derivation of these results is similar to that used in Ref.~[\onlinecite{Kane:prl97}]
to discuss tunneling into a finite carbon nanotube.) 

The amplitude for momentum-resolved tunneling is determined by the
Fourier transform of (\ref{Gfin}) with respect to the difference
variable $(x-x')$.  Thus we recover the $k$-dependence $A(k,\bar u)$,
given by (\ref{Adef}) with $\bar u = \bar u (L/v_c)$.

We see that in the limit of large $N$ there are two qualitatively
different features to be expected in the tunneling between parallel
quantum wires in the free-spin (spin incoherent) regime compared to
the ground state (spin coherent) tunneling regime: 
(1) The momentum structure is no longer
sharply peaked at $k=\pm k_F$ at zero bias with the peaks splitting at the spin/charge velocity slopes at a finite bias, but depends on the
relative size of $\bar u$ and the mean electron spacing $a$. When
$\bar u \gg a$, the momentum distribution has a single broad peak centered at $k=0$, while
when $\bar u \ll  a$, the momentum distribution has a double lobed
structure peaked at $k=\pm \tilde k_F$ at zero bias with the peak position shifting at the charge-velocity slope at a finite bias.  Note, also, that $ \tilde k_F$ is twice as large as $k_F$ for unpolarized electrons at the same density.  (2) Depending on
the interaction parameter $g$ of the charge sector, it is possible to
have a diverging tunneling density of states as the energy is lowered.
This behavior contrasts with the ubiquitous power law suppression of the 
spin-coherent Luttinger Liquid regime.  

\subsection{Effects of non-zero Zeeman  field}
\label{nzf}

Our results for the free-spin regime can be generalized to the situation where the Zeeman energy 
 $E_Z=g_s\mu_B B$ is comparable to $k_B T$, but large compared to the exchange energy $J$.  In this case the spins will be partially polarized, but uncorrelated, and we can use essentially the same analysis as before. The primary modification is that the probability $p_m$ for  $|m|+1$ consecutive spins that are all spin up or all spin down, respectively, which appears in Eq. (\ref{eq:G_Bert}), 
is now given by 
\begin{equation}
p_m = w_+^{|m|+1} + w_-^{|m|+1}\;,  
\end{equation}
where $w_+ = 1- w_- = (1 + e^{- E_Z / k_B T})^{-1}$ is the probability for finding a spin aligned parallel to the field. 

To implement this change, the factors $2^{-|j|}$ in (\ref{eq:G_0}) and (\ref{fdef}) 
must be replaced by $p_j$ .  The function $A(k,\bar u )$, which describes the  momentum dependence of the tunneling rate, is now given by
\begin{equation}
\label{AdefB}
A(k, \bar u) = e^{-k^2 \bar u^2 /2} (F_+ + F_-) \;,
\end{equation}
\begin{equation}
\label{eq:F}
F_{\pm} \equiv \frac{w_{\pm}(1 -w_{\pm}^2)}{1 + w_{\pm}^2 + 2 w_{\pm} \cos(ka)}\;.
\end{equation}

It may be readily seen that when the degree of polarization is increased, the peaks of $A(k,\bar u)$  at $k = \pm {\tilde{k}}_F$ become higher and narrower, while the weight decreases elsewhere.(The higher momentum peaks are exponentially supressed by the prefactor in Eq.~(\ref{AdefB}).) If $ka$ is fixed at any value other than an odd multiple of $\pi$, then the value of $A(k, \bar u)$ will vanish in the limit $w_+ \to 1$.  By contrast, if $ka$ is equal to an odd multiple of $\pi$, the value of 
 $A(k, \bar u)$ will diverge in this limit, proportional to $w_-^{-1}$. The function $A(k, \bar u)$ is plotted in Fig.~\ref{fig:A_k_u_x} for several values
of $x=E_Z/k_BT$ with $\bar u=0.6 a$. Compare with Fig.~\ref{fig:smallx_FT_u}.

\begin{figure}
\includegraphics[width=1.0\linewidth,clip=]{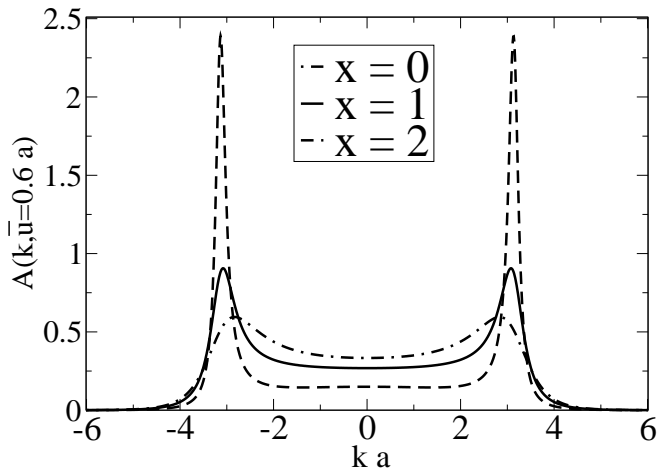}
\caption{\label{fig:A_k_u_x} Polarization dependence of the spectral function given by Eqs.~(\ref{AdefB}) and (\ref{eq:F}).
Here $x=E_Z/k_BT$ with $E_Z$ the Zeeman energy and $T$ the temperature. We have used $\bar u =0.6 a$ for each curve. (See Fig.~\ref{fig:smallx_FT_u} for the dependence on $\bar u$ at $E_Z=0$.) As the
polarization increases, the peaks at $\tilde k_F=\pi/a$ become narrower and 
higher as described in the text and the weight between the peaks decreases.}
\end{figure}

To estimate the amplitude for resonant tunneling into a single point near the center of the wire, we can use (\ref{Gfin}), with $x=x'=L/2$ .  The function $f$ is to  be evaluated using (\ref{fdef}), with the factor $2^{-|j|}$ replaced by $p_j$, and $\bar u$ given by (\ref{uL}) .  For any value of the polarization less than unity, the value of $f(z=0, \bar u)$ is finite in the limit  $\bar u \to \infty$, so we find 
\begin{equation}
\mathcal{G}(x,x)  \sim  L^{-(1+\tilde{\alpha}) } \ln^{-1/2}(L/a)\; , 
\end{equation}
for $L\to \infty$ with  ${\tilde{\alpha}}$ given by (\ref{alfs}). 
On the other hand, for a fully polarized system, with $w_+ = 1$, one finds that $f(z=0, \bar u)$ vanishes like $\bar u L^{-g} $ for $L \to \infty$ and $\bar u$ given by (\ref{uL}).  One then finds 
\begin{equation}
\mathcal{G}(x,x) \sim 1/L^{(1+  \alpha_p)} \;,
\end {equation}
with $\alpha_p = g + (4 g)^{-1} - 1 $ . This exponent is the same as the standard tunneling exponent one would obtain for spinless fermions having the same interaction and density, if one takes into account the difference in the definitions of $g$ for the two cases.  (Here,  we have defined $g$  as  $v_F / v_c$, where $v_F$, the Fermi velocity for non-interacting  spinless fermions, is half of the Fermi velocity for spinless fermions at the same density.)

\section{Comparison with experiments}
\label{sec:experiments}

Results for the momentum-dependence of the tunneling conductance,
obtained in the present paper, show various similarities and
differences with the experimental results of Steinberg et
al.\cite{Steinberg:sci04,Steinberg:sci05} in the localized regime
(that is, the regime in which the tunneling exhibits broad structure
in momentum space and hence local structure in position space).  In
general, experimental results at the first Coulomb blockade peak,
associated with the first electron to tunnel into the localized
region, show a single peak, centered at $k=0$, as expected from our
analysis.  However, the spectrum for the second and subsequent
electrons typically show two peaks, at positions $\approx \pm k_m$
that increase with increasing $N$.  The intensity between the two
peaks is not small, being perhaps half the intensity at the maxima,
and one does not observe the zeroes of the intensity that we find for
ground-state tunneling in a symmetric well.

Qualitatively, the experiments are most consistent with our results
for the free-spin regime, i.e., the results one would expect when
$k_BT$ is large compared to $J$ but small compared to the energy for
charge excitations.  According to the results of Subsection \ref{see}
above, $J$ should indeed be smaller than the experimental temperature
of 0.25~K, for densities less than about 6 electrons per $\mu$m. The
confinement length $L$ in the experiments is not precisely known, and
it may well change as additional electrons are added, but if one
estimates $L$ in the range of 0.5 to 1~$\mu$m, the free-spin regime
would hold for values of $N$ less than perhaps 3 to 6. On the other
hand, there is evidence that for the largest values of $N$ in the
localized regime, the density reaches 15 electrons per $\mu$m, for
which we estimate a value of $J$ much larger than $k_BT$.

An added factor in the experiments is that the Zeeman energy is not
negligible compared to $k_BT$.  For a typical applied field of 2~T,
the Zeeman energy is about 0.6~K, more than twice the quoted
temperature.  If the Zeeman energy is comparable to, but smaller than
$J$, the system will be partially polarized even at $T=0$.  For $J\sim
E_Z>k_B T>J/N$, typically a number of different spin states will be
occupied.  Theoretical expectations for the case with a sizable Zeeman
energy were discussed briefly in Subsection \ref{nzf} above, for the
free spin regime, and they may be qualitatively consistent with the
experiments.

We recall that in the free-spin regime (provided that $\bar u < 0.75
a$), or in the fully-polarized regime, the position of the intensity
maximum should occur at $k_m \approx 2 k_F$, where $k_F$ is the Fermi
wavevector for unpolarized electrons at the same density. There is no
experimental indication of this when the upper wire is in the
delocalized (that is, the Luttinger Liquid regime) regime.  Thus if
the electrons in the localized regime (see first paragraph of this
section above) form either a free-spin or a spin-polarized lattice,
there should be an increase in the value of $k_m$ as one enters the
localized regime.  It is not clear whether such an effect is observed
in the experiments. The tendency for an increase in $k_m$ may be
counteracted if there is a discontinuous decrease in electron density
upon entering the localized regime.

The situation where $k_BT$ and $E_Z$ are both large compared to $J$
was discussed for the simplest case, $N=2$, in Subsection \ref{mss},
and illustrated in Fig.~\ref{fig:FT12}. Under the experimental
conditions where $E_Z$ is large compared to $k_BT$, the momentum
dependence of the conductivity should be close to that of the pure
triplet state, and should have near zero intensity at $k=0$, if the
confining well is symmetric.  This contrasts with the experimental
data for $N=2$, where the intensity at $k=0$ is a substantial fraction
of the maximum intensity.  The discrepancy might be explained if the
confining well is sufficiently asymmetric, or if the electron
temperature in the upper wire is significantly higher than the lattice
temperature.

The calculations in Section V above used a model with sharp
confinement at the ends of the wire and a flat potential within. We
presented exact numerical diagonalizations for two and four electrons. A
model with soft confinement, and higher density near the center of the
wire, would lead to more intensity at $|k|<k_m$, and a faster fall-off
of intensity for $|k|> k_m$, than one obtains in the case of sharp
confinement.\cite{yaro02,yaro03}

Finally, we repeat that the reason for the apparent formation of a
barrier between the low-density central section of the upper wire and
the high density outer regions is not well understood.

\section{Conclusions} 
\label{sec:conclusions}

In this paper we have studied momentum resolved tunneling into a short
quantum wire.  We have focused on the situation where the number of
electrons, $N$, is not too large and the density of electrons is low
enough to approach the Wigner crystal limit.  We began by reviewing
some general theorems which dictate basic properties of the momentum
dependence.  We note that the matrix element for tunneling between
ground states for $N$ and $N-1$ electrons can be expressed in terms of
the Fourier transform of a real function $\Psi_{\rm{eff}}(x)$, which
is the matrix element of $\psi_\sigma (x)$, the electron annihilation
operator at point $x$, between the two states. For non-interacting
electrons, $\Psi_{\rm{eff}}(x)$, is the wavefunction of the highest
filled level in the $N$-electron system; more generally,
$\Psi_{\rm{eff}}(x)$ is a quasi-wavefunction, whose magnitude is
reduced by many-body correlations, but whose spatial dependence is
qualitatively similar to the non-interacting case.

Using exact diagonalization, we computed the ground-state density, the
density-density correlation function, and the effective exchange
constant, $J$, for $N \leq 4$.  We include screening effects from the
adjacent, higher density wire.  In the limit of large $N$ we applied
bosonization techniques to study the momentum resolved tunneling, both
at $T=0$, and in the free-spin limit, of spin energies much less than
temperature and the Fermi energy, $J \ll k_BT \ll E_F$.  Whereas for
$k_BT \ll J, E_F$, the momentum dependence has narrow peaks centered
at $k=\pm k_F$, in the free-spin regime, the spectral function can
have broad peaks centered at $k=\pm 2 k_F$. Comparisons of our
theoretical predictions with experimental results obtained by
Steinberg et al.\cite{Steinberg:sci04,Steinberg:sci05} show qualitative agreement on
many features, but some aspects are still not understood, and further
work is needed.

While we were finishing this manuscript, we became aware of
independent unpublished work by E. J. Mueller, based on the
experiments that motivated our study.\cite{Mueller:cm04} Mueller's
work employs, primarily, a Hartree-Fock approximation, and is largely
orthogonal to the work reported here.

\acknowledgments

We thank Ophir Auslaender, Leon Balents, Leonid Glazman, Walter
Hofstetter, Karyn Le Hur, Hadar Steinberg, and Amir Yacoby for
stimulating discussions.  G.A.F. thanks Misha Fogler and the physics
department at UCSD, where part of this work was done, for hospitality.
This work was supported by NSF PHY99-07949, DMR02-33773, the Packard
Foundation, and the Harvard Society of Fellows.


\begin{thebibliography}{55}
\expandafter\ifx\csname natexlab\endcsname\relax\def\natexlab#1{#1}\fi
\expandafter\ifx\csname bibnamefont\endcsname\relax
  \def\bibnamefont#1{#1}\fi
\expandafter\ifx\csname bibfnamefont\endcsname\relax
  \def\bibfnamefont#1{#1}\fi
\expandafter\ifx\csname citenamefont\endcsname\relax
  \def\citenamefont#1{#1}\fi
\expandafter\ifx\csname url\endcsname\relax
  \def\url#1{\texttt{#1}}\fi
\expandafter\ifx\csname urlprefix\endcsname\relax\def\urlprefix{URL }\fi
\providecommand{\bibinfo}[2]{#2}
\providecommand{\eprint}[2][]{\url{#2}}

\bibitem[{\citenamefont{Black et~al.}(1996)\citenamefont{Black, Ralph, and
  Tinkham}}]{Black:prl96}
\bibinfo{author}{\bibfnamefont{C.~T.} \bibnamefont{Black}},
  \bibinfo{author}{\bibfnamefont{D.~C.} \bibnamefont{Ralph}}, \bibnamefont{and}
  \bibinfo{author}{\bibfnamefont{M.}~\bibnamefont{Tinkham}},
  \bibinfo{journal}{Phys. Rev. Lett.} \textbf{\bibinfo{volume}{76}},
  \bibinfo{pages}{688} (\bibinfo{year}{1996}).

\bibitem[{\citenamefont{Ralph et~al.}(1997)\citenamefont{Ralph, Black, and
  Tinkham}}]{Ralph:prl96}
\bibinfo{author}{\bibfnamefont{D.~C.} \bibnamefont{Ralph}},
  \bibinfo{author}{\bibfnamefont{C.~T.} \bibnamefont{Black}}, \bibnamefont{and}
  \bibinfo{author}{\bibfnamefont{M.}~\bibnamefont{Tinkham}},
  \bibinfo{journal}{Phys. Rev. Lett.} \textbf{\bibinfo{volume}{78}},
  \bibinfo{pages}{4087} (\bibinfo{year}{1997}).

\bibitem[{\citenamefont{Goldhaber-Gordon
  et~al.}(1998)\citenamefont{Goldhaber-Gordon, Shtrikman, Mahalu,
  Abusch-Magder, Meirav, and Kastner}}]{Goldhaber:nat98}
\bibinfo{author}{\bibfnamefont{D.}~\bibnamefont{Goldhaber-Gordon}},
  \bibinfo{author}{\bibfnamefont{H.}~\bibnamefont{Shtrikman}},
  \bibinfo{author}{\bibfnamefont{D.}~\bibnamefont{Mahalu}},
  \bibinfo{author}{\bibfnamefont{D.}~\bibnamefont{Abusch-Magder}},
  \bibinfo{author}{\bibfnamefont{U.}~\bibnamefont{Meirav}}, \bibnamefont{and}
  \bibinfo{author}{\bibfnamefont{M.~A.} \bibnamefont{Kastner}},
  \bibinfo{journal}{Nature} \textbf{\bibinfo{volume}{{\bf 391}}},
  \bibinfo{pages}{156} (\bibinfo{year}{1998}).

\bibitem[{\citenamefont{Manoharan et~al.}(2000)\citenamefont{Manoharan, Lutz,
  and Eigler}}]{Manoharan:nat00}
\bibinfo{author}{\bibfnamefont{H.~C.} \bibnamefont{Manoharan}},
  \bibinfo{author}{\bibfnamefont{C.~P.} \bibnamefont{Lutz}}, \bibnamefont{and}
  \bibinfo{author}{\bibfnamefont{D.~M.} \bibnamefont{Eigler}},
  \bibinfo{journal}{Nature} \textbf{\bibinfo{volume}{{\bf 403}}},
  \bibinfo{pages}{512} (\bibinfo{year}{2000}).

\bibitem[{\citenamefont{Nygard et~al.}(2000)\citenamefont{Nygard, Cobden, and
  Lindelof}}]{Nygard:nat00}
\bibinfo{author}{\bibfnamefont{J.}~\bibnamefont{Nygard}},
  \bibinfo{author}{\bibfnamefont{D.~H.} \bibnamefont{Cobden}},
  \bibnamefont{and} \bibinfo{author}{\bibfnamefont{P.~E.}
  \bibnamefont{Lindelof}}, \bibinfo{journal}{Nature}
  \textbf{\bibinfo{volume}{{\bf 408}}}, \bibinfo{pages}{342}
  (\bibinfo{year}{2000}).

\bibitem[{\citenamefont{Bockrath et~al.}(1999)\citenamefont{Bockrath, Cobden,
  Lu, Rinzler, Smalley, Balents, and McEuen}}]{Bockrath:nat99}
\bibinfo{author}{\bibfnamefont{M.}~\bibnamefont{Bockrath}},
  \bibinfo{author}{\bibfnamefont{D.~H.} \bibnamefont{Cobden}},
  \bibinfo{author}{\bibfnamefont{J.}~\bibnamefont{Lu}},
  \bibinfo{author}{\bibfnamefont{A.~G.} \bibnamefont{Rinzler}},
  \bibinfo{author}{\bibfnamefont{R.~E.} \bibnamefont{Smalley}},
  \bibinfo{author}{\bibfnamefont{L.}~\bibnamefont{Balents}}, \bibnamefont{and}
  \bibinfo{author}{\bibfnamefont{P.~L.} \bibnamefont{McEuen}},
  \bibinfo{journal}{Nature} \textbf{\bibinfo{volume}{397}},
  \bibinfo{pages}{598} (\bibinfo{year}{1999}).

\bibitem[{\citenamefont{Auslaender et~al.}(2002)\citenamefont{Auslaender,
  Yacoby, de~Picciotto, Baldwin, Pfeiffer, and West}}]{yacoby:sci02}
\bibinfo{author}{\bibfnamefont{O.}~\bibnamefont{Auslaender}},
  \bibinfo{author}{\bibfnamefont{A.}~\bibnamefont{Yacoby}},
  \bibinfo{author}{\bibfnamefont{R.}~\bibnamefont{de~Picciotto}},
  \bibinfo{author}{\bibfnamefont{K.~W.} \bibnamefont{Baldwin}},
  \bibinfo{author}{\bibfnamefont{L.~N.} \bibnamefont{Pfeiffer}},
  \bibnamefont{and} \bibinfo{author}{\bibfnamefont{K.~W.} \bibnamefont{West}},
  \bibinfo{journal}{Science} \textbf{\bibinfo{volume}{295}},
  \bibinfo{pages}{825} (\bibinfo{year}{2002}).

\bibitem[{\citenamefont{Tserkovnyak et~al.}(2002)\citenamefont{Tserkovnyak,
  Halperin, Auslaender, and Yacoby}}]{yaro02}
\bibinfo{author}{\bibfnamefont{Y.}~\bibnamefont{Tserkovnyak}},
  \bibinfo{author}{\bibfnamefont{B.~I.} \bibnamefont{Halperin}},
  \bibinfo{author}{\bibfnamefont{O.~M.} \bibnamefont{Auslaender}},
  \bibnamefont{and} \bibinfo{author}{\bibfnamefont{A.}~\bibnamefont{Yacoby}},
  \bibinfo{journal}{Phys. Rev. Lett.} \textbf{\bibinfo{volume}{89}},
  \bibinfo{pages}{136805} (\bibinfo{year}{2002}).

\bibitem[{\citenamefont{Ishii et~al.}(2003)\citenamefont{Ishii, Kataura,
  Shiozawa, Yoshioka, Otsubo, Takayama, Miyahara, Suzuki, Achiba, Nakatake
  et~al.}}]{ishii:nat03}
\bibinfo{author}{\bibfnamefont{H.}~\bibnamefont{Ishii}},
  \bibinfo{author}{\bibfnamefont{H.}~\bibnamefont{Kataura}},
  \bibinfo{author}{\bibfnamefont{H.}~\bibnamefont{Shiozawa}},
  \bibinfo{author}{\bibfnamefont{H.}~\bibnamefont{Yoshioka}},
  \bibinfo{author}{\bibfnamefont{H.}~\bibnamefont{Otsubo}},
  \bibinfo{author}{\bibfnamefont{Y.}~\bibnamefont{Takayama}},
  \bibinfo{author}{\bibfnamefont{T.}~\bibnamefont{Miyahara}},
  \bibinfo{author}{\bibfnamefont{S.}~\bibnamefont{Suzuki}},
  \bibinfo{author}{\bibfnamefont{Y.}~\bibnamefont{Achiba}},
  \bibinfo{author}{\bibfnamefont{M.}~\bibnamefont{Nakatake}},
  \bibnamefont{et~al.}, \bibinfo{journal}{Nature}
  \textbf{\bibinfo{volume}{426}}, \bibinfo{pages}{540} (\bibinfo{year}{2003}).

\bibitem[{\citenamefont{Auslaender et~al.}(2000)\citenamefont{Auslaender,
  Yacoby, de~Picciotto, Baldwin, Pfeiffer, and West}}]{Auslaender:prl00}
\bibinfo{author}{\bibfnamefont{O.~M.} \bibnamefont{Auslaender}},
  \bibinfo{author}{\bibfnamefont{A.}~\bibnamefont{Yacoby}},
  \bibinfo{author}{\bibfnamefont{R.}~\bibnamefont{de~Picciotto}},
  \bibinfo{author}{\bibfnamefont{K.~W.} \bibnamefont{Baldwin}},
  \bibinfo{author}{\bibfnamefont{L.~N.} \bibnamefont{Pfeiffer}},
  \bibnamefont{and} \bibinfo{author}{\bibfnamefont{K.~W.} \bibnamefont{West}},
  \bibinfo{journal}{Phys. Rev. Lett.} \textbf{\bibinfo{volume}{84}},
  \bibinfo{pages}{1764} (\bibinfo{year}{2000}).

\bibitem[{\citenamefont{Pfeiffer et~al.}(1997)\citenamefont{Pfeiffer, Yacoby,
  Stormer, Baldwin, Hasen, Pinczuk, Wegscheider, and West}}]{Pfeiffer:mej97}
\bibinfo{author}{\bibfnamefont{L.}~\bibnamefont{Pfeiffer}},
  \bibinfo{author}{\bibfnamefont{A.}~\bibnamefont{Yacoby}},
  \bibinfo{author}{\bibfnamefont{H.}~\bibnamefont{Stormer}},
  \bibinfo{author}{\bibfnamefont{K.}~\bibnamefont{Baldwin}},
  \bibinfo{author}{\bibfnamefont{J.}~\bibnamefont{Hasen}},
  \bibinfo{author}{\bibfnamefont{A.}~\bibnamefont{Pinczuk}},
  \bibinfo{author}{\bibfnamefont{W.}~\bibnamefont{Wegscheider}},
  \bibnamefont{and} \bibinfo{author}{\bibfnamefont{K.}~\bibnamefont{West}},
  \bibinfo{journal}{Microelectron. J.} \textbf{\bibinfo{volume}{28}},
  \bibinfo{pages}{817} (\bibinfo{year}{1997}).

\bibitem[{\citenamefont{Tserkovnyak et~al.}(2003)\citenamefont{Tserkovnyak,
  Halperin, Auslaender, and Yacoby}}]{yaro03}
\bibinfo{author}{\bibfnamefont{Y.}~\bibnamefont{Tserkovnyak}},
  \bibinfo{author}{\bibfnamefont{B.~I.} \bibnamefont{Halperin}},
  \bibinfo{author}{\bibfnamefont{O.~M.} \bibnamefont{Auslaender}},
  \bibnamefont{and} \bibinfo{author}{\bibfnamefont{A.}~\bibnamefont{Yacoby}},
  \bibinfo{journal}{Phys. Rev. B} \textbf{\bibinfo{volume}{68}},
  \bibinfo{pages}{125312} (\bibinfo{year}{2003}).

\bibitem[{Kak()}]{Kakashvili:prl03}
\bibinfo{note}{A similar geometry to that shown in Fig.1 was considered in
  Kakashvili and Johannesson, PRL {\bf 91,} 186403 (2003). However, in
  Kakashvili and Johannesson tunneling was allowed only at a point, so no
  momentum information on the tunneling could be obtained, in contrast to the
  present work.}

\bibitem[{\citenamefont{Carpentier et~al.}(2002)\citenamefont{Carpentier, Peca,
  and Balents}}]{Carpentier:prb02}
\bibinfo{author}{\bibfnamefont{D.}~\bibnamefont{Carpentier}},
  \bibinfo{author}{\bibfnamefont{C.}~\bibnamefont{Peca}}, \bibnamefont{and}
  \bibinfo{author}{\bibfnamefont{L.}~\bibnamefont{Balents}},
  \bibinfo{journal}{Phys. Rev. B} \textbf{\bibinfo{volume}{66}},
  \bibinfo{pages}{153304} (\bibinfo{year}{2002}).

\bibitem[{\citenamefont{Z\"ulicke and Governale}(2002)}]{Zulicke:prb02}
\bibinfo{author}{\bibfnamefont{U.}~\bibnamefont{Z\"ulicke}} \bibnamefont{and}
  \bibinfo{author}{\bibfnamefont{M.}~\bibnamefont{Governale}},
  \bibinfo{journal}{Phys. Rev. B} \textbf{\bibinfo{volume}{65}},
  \bibinfo{pages}{205304} (\bibinfo{year}{2002}).

\bibitem[{\citenamefont{Boese et~al.}(2001)\citenamefont{Boese, Governale,
  Rosch, and Z\"ulike}}]{Boese:prb01}
\bibinfo{author}{\bibfnamefont{D.}~\bibnamefont{Boese}},
  \bibinfo{author}{\bibfnamefont{M.}~\bibnamefont{Governale}},
  \bibinfo{author}{\bibfnamefont{A.}~\bibnamefont{Rosch}}, \bibnamefont{and}
  \bibinfo{author}{\bibfnamefont{U.}~\bibnamefont{Z\"ulike}},
  \bibinfo{journal}{Phys. Rev. B} \textbf{\bibinfo{volume}{64}},
  \bibinfo{pages}{085315} (\bibinfo{year}{2001}).

\bibitem[{\citenamefont{Atland et~al.}(1999)\citenamefont{Atland, Barnes,
  Hekking, and Schofield}}]{Atland:prl99}
\bibinfo{author}{\bibfnamefont{A.}~\bibnamefont{Atland}},
  \bibinfo{author}{\bibfnamefont{C.~H.~W.} \bibnamefont{Barnes}},
  \bibinfo{author}{\bibfnamefont{F.~W.~J.} \bibnamefont{Hekking}},
  \bibnamefont{and} \bibinfo{author}{\bibfnamefont{A.~J.}
  \bibnamefont{Schofield}}, \bibinfo{journal}{Phys. Rev. Lett.}
  \textbf{\bibinfo{volume}{83}}, \bibinfo{pages}{1203} (\bibinfo{year}{1999}).

\bibitem[{\citenamefont{Governale et~al.}(2000)\citenamefont{Governale,
  Grifoni, and Sch\"on}}]{Governale:prb00}
\bibinfo{author}{\bibfnamefont{M.}~\bibnamefont{Governale}},
  \bibinfo{author}{\bibfnamefont{M.}~\bibnamefont{Grifoni}}, \bibnamefont{and}
  \bibinfo{author}{\bibfnamefont{G.}~\bibnamefont{Sch\"on}},
  \bibinfo{journal}{Phys. Rev. B} \textbf{\bibinfo{volume}{62}},
  \bibinfo{pages}{15996} (\bibinfo{year}{2000}).

\bibitem[{\citenamefont{Eggert et~al.}(1996)\citenamefont{Eggert, Johannesson,
  and Mattsson}}]{Eggert:prl96}
\bibinfo{author}{\bibfnamefont{S.}~\bibnamefont{Eggert}},
  \bibinfo{author}{\bibfnamefont{H.}~\bibnamefont{Johannesson}},
  \bibnamefont{and} \bibinfo{author}{\bibfnamefont{A.~H.}
  \bibnamefont{Mattsson}}, \bibinfo{journal}{Phys. Rev. Lett.}
  \textbf{\bibinfo{volume}{76}}, \bibinfo{pages}{1505} (\bibinfo{year}{1996}).

\bibitem[{\citenamefont{Mattsson et~al.}(1997)\citenamefont{Mattsson, Eggert,
  and Johannesson}}]{Mattsson:prb97}
\bibinfo{author}{\bibfnamefont{A.~H.} \bibnamefont{Mattsson}},
  \bibinfo{author}{\bibfnamefont{S.}~\bibnamefont{Eggert}}, \bibnamefont{and}
  \bibinfo{author}{\bibfnamefont{H.}~\bibnamefont{Johannesson}},
  \bibinfo{journal}{Phys. Rev. B} \textbf{\bibinfo{volume}{56}},
  \bibinfo{pages}{15615} (\bibinfo{year}{1997}).

\bibitem[{\citenamefont{Fabrizio and Gogolin}(1995)}]{Fabrizio:prb95}
\bibinfo{author}{\bibfnamefont{M.}~\bibnamefont{Fabrizio}} \bibnamefont{and}
  \bibinfo{author}{\bibfnamefont{A.~O.} \bibnamefont{Gogolin}},
  \bibinfo{journal}{Phys. Rev. B} \textbf{\bibinfo{volume}{51}},
  \bibinfo{pages}{17827} (\bibinfo{year}{1995}).

\bibitem[{\citenamefont{Kleimann et~al.}(2002)\citenamefont{Kleimann,
  Cavaliere, Sassetti, and Kramer}}]{Kleimann:prb02}
\bibinfo{author}{\bibfnamefont{T.}~\bibnamefont{Kleimann}},
  \bibinfo{author}{\bibfnamefont{F.}~\bibnamefont{Cavaliere}},
  \bibinfo{author}{\bibfnamefont{M.}~\bibnamefont{Sassetti}}, \bibnamefont{and}
  \bibinfo{author}{\bibfnamefont{B.}~\bibnamefont{Kramer}},
  \bibinfo{journal}{Phys. Rev. B} \textbf{\bibinfo{volume}{66}},
  \bibinfo{pages}{165311} (\bibinfo{year}{2002}).

\bibitem[{\citenamefont{Kleimann et~al.}(2000)\citenamefont{Kleimann, Sassetti,
  Kramer, and Yacoby}}]{Kleimann:prb00}
\bibinfo{author}{\bibfnamefont{T.}~\bibnamefont{Kleimann}},
  \bibinfo{author}{\bibfnamefont{M.}~\bibnamefont{Sassetti}},
  \bibinfo{author}{\bibfnamefont{B.}~\bibnamefont{Kramer}}, \bibnamefont{and}
  \bibinfo{author}{\bibfnamefont{A.}~\bibnamefont{Yacoby}},
  \bibinfo{journal}{Phys. Rev. B} \textbf{\bibinfo{volume}{62}},
  \bibinfo{pages}{8144} (\bibinfo{year}{2000}).

\bibitem[{Ste()}]{Steinberg:sci04}
\bibinfo{note}{H. Steinberg, O. M. Auslaender, A. Yacoby, J. Qian, G. A. Fiete,
  Y. Tserkovnyak, B. I. Halperin, R. de Picciotto, K. W. Baldwin, L. N.
  Pfeiffer and K. W. West (unpublished).}

\bibitem[{\citenamefont{Auslaender et~al.}(2005)\citenamefont{Auslaender,
  Steinberg, Yacoby, Tserkovnyak, Halperin, Baldwin, Pfeiffer, and
  West}}]{Steinberg:sci05}
\bibinfo{author}{\bibfnamefont{O.~M.} \bibnamefont{Auslaender}},
  \bibinfo{author}{\bibfnamefont{H.}~\bibnamefont{Steinberg}},
  \bibinfo{author}{\bibfnamefont{A.}~\bibnamefont{Yacoby}},
  \bibinfo{author}{\bibfnamefont{Y.}~\bibnamefont{Tserkovnyak}},
  \bibinfo{author}{\bibfnamefont{B.~I.} \bibnamefont{Halperin}},
  \bibinfo{author}{\bibfnamefont{K.~W.} \bibnamefont{Baldwin}},
  \bibinfo{author}{\bibfnamefont{L.~N.} \bibnamefont{Pfeiffer}},
  \bibnamefont{and} \bibinfo{author}{\bibfnamefont{K.~W.} \bibnamefont{West}},
  \bibinfo{journal}{Science} \textbf{\bibinfo{volume}{308}},
  \bibinfo{pages}{88} (\bibinfo{year}{2005}).

\bibitem[{qia()}]{qian03}
\bibinfo{note}{J. Qian, G. Zar\'and, W. Hofstetter and B.~I. Halperin
  (unpublished).}

\bibitem[{\citenamefont{Hirose et~al.}(2003)\citenamefont{Hirose, Meir, and
  Wingreen}}]{hirose03}
\bibinfo{author}{\bibfnamefont{K.}~\bibnamefont{Hirose}},
  \bibinfo{author}{\bibfnamefont{Y.}~\bibnamefont{Meir}}, \bibnamefont{and}
  \bibinfo{author}{\bibfnamefont{N.~S.} \bibnamefont{Wingreen}},
  \bibinfo{journal}{Phys. Rev. Lett.} \textbf{\bibinfo{volume}{90}},
  \bibinfo{pages}{026804} (\bibinfo{year}{2003}).

\bibitem[{\citenamefont{Tanatar et~al.}(1998)\citenamefont{Tanatar, Al-Hayek,
  and Tomak}}]{tanatar98}
\bibinfo{author}{\bibfnamefont{B.}~\bibnamefont{Tanatar}},
  \bibinfo{author}{\bibfnamefont{I.}~\bibnamefont{Al-Hayek}}, \bibnamefont{and}
  \bibinfo{author}{\bibfnamefont{M.}~\bibnamefont{Tomak}},
  \bibinfo{journal}{Phys. Rev. B} \textbf{\bibinfo{volume}{58}},
  \bibinfo{pages}{9886} (\bibinfo{year}{1998}).

\bibitem[{\citenamefont{Glazman et~al.}(1992)\citenamefont{Glazman, Ruzin, and
  Shklovskii}}]{Glazman:prb92}
\bibinfo{author}{\bibfnamefont{L.~I.} \bibnamefont{Glazman}},
  \bibinfo{author}{\bibfnamefont{I.~M.} \bibnamefont{Ruzin}}, \bibnamefont{and}
  \bibinfo{author}{\bibfnamefont{B.~I.} \bibnamefont{Shklovskii}},
  \bibinfo{journal}{Phys. Rev. B} \textbf{\bibinfo{volume}{45}},
  \bibinfo{pages}{8454} (\bibinfo{year}{1992}).

\bibitem[{\citenamefont{H\"ausler et~al.}(2002)\citenamefont{H\"ausler, Kecke,
  and MacDonald}}]{Hausler:prb02}
\bibinfo{author}{\bibfnamefont{W.}~\bibnamefont{H\"ausler}},
  \bibinfo{author}{\bibfnamefont{L.}~\bibnamefont{Kecke}}, \bibnamefont{and}
  \bibinfo{author}{\bibfnamefont{A.~H.} \bibnamefont{MacDonald}},
  \bibinfo{journal}{Phys. Rev. B} \textbf{\bibinfo{volume}{65}},
  \bibinfo{pages}{085104} (\bibinfo{year}{2002}).

\bibitem[{\citenamefont{Matveev}(2004{\natexlab{a}})}]{Matveev:prl04}
\bibinfo{author}{\bibfnamefont{K.~A.} \bibnamefont{Matveev}},
  \bibinfo{journal}{Phys. Rev. Lett.} \textbf{\bibinfo{volume}{92}},
  \bibinfo{pages}{106801} (\bibinfo{year}{2004}{\natexlab{a}}).

\bibitem[{\citenamefont{Matveev}(2004{\natexlab{b}})}]{Matveev:prb04}
\bibinfo{author}{\bibfnamefont{K.~A.} \bibnamefont{Matveev}},
  \bibinfo{journal}{Phys. Rev. B} \textbf{\bibinfo{volume}{70}},
  \bibinfo{pages}{245319} (\bibinfo{year}{2004}{\natexlab{b}}).

\bibitem[{\citenamefont{Schulz}(1993)}]{Schulz:prl93}
\bibinfo{author}{\bibfnamefont{H.~J.} \bibnamefont{Schulz}},
  \bibinfo{journal}{Phys. Rev. Lett.} \textbf{\bibinfo{volume}{71}},
  \bibinfo{pages}{1864} (\bibinfo{year}{1993}).

\bibitem[{\citenamefont{Lieb and Mattis}(1962)}]{lieb62}
\bibinfo{author}{\bibfnamefont{E.}~\bibnamefont{Lieb}} \bibnamefont{and}
  \bibinfo{author}{\bibfnamefont{D.}~\bibnamefont{Mattis}},
  \bibinfo{journal}{Phys. Rev.} \textbf{\bibinfo{volume}{125}},
  \bibinfo{pages}{164} (\bibinfo{year}{1962}).

\bibitem[{Lie()}]{Lieb:footnote}
\bibinfo{note}{The theorem of Ref.~[\onlinecite{lieb62}] holds for the
  one-dimensional Schroedinger equation with spin-independent many-body
  interactions that are arbitrary functions of position.}

\bibitem[{\citenamefont{Dagotto}(1994)}]{Dagotto:rmp94}
\bibinfo{author}{\bibfnamefont{E.}~\bibnamefont{Dagotto}},
  \bibinfo{journal}{Rev. Mod. Phys.} \textbf{\bibinfo{volume}{66}},
  \bibinfo{pages}{763} (\bibinfo{year}{1994}).

\bibitem[{\citenamefont{Cullum and Willoughby}(1985)}]{lanczos}
\bibinfo{author}{\bibfnamefont{J.~K.} \bibnamefont{Cullum}} \bibnamefont{and}
  \bibinfo{author}{\bibfnamefont{R.~A.} \bibnamefont{Willoughby}},
  \emph{\bibinfo{title}{Lanczos Method for Large Symmetric Eigenvalue
  Computation, Vol. 1}} (\bibinfo{publisher}{Birkhauser Boston Inc.},
  \bibinfo{year}{1985}).

\bibitem[{\citenamefont{H\"auseler and Kramer}(1993)}]{Hausler:prb93}
\bibinfo{author}{\bibfnamefont{W.}~\bibnamefont{H\"auseler}} \bibnamefont{and}
  \bibinfo{author}{\bibfnamefont{B.}~\bibnamefont{Kramer}},
  \bibinfo{journal}{Phys. Rev. B} \textbf{\bibinfo{volume}{47}},
  \bibinfo{pages}{16353} (\bibinfo{year}{1993}). See also W. H\"ausler, 
Annalen der Physik {\bf 5}, 401 (1996).

\bibitem[{\citenamefont{Jauregui et~al.}(1993)\citenamefont{Jauregui,
  H\"auseler, and Kramer}}]{Jauregui:epl96}
\bibinfo{author}{\bibfnamefont{K.}~\bibnamefont{Jauregui}},
  \bibinfo{author}{\bibfnamefont{W.}~\bibnamefont{H\"auseler}},
  \bibnamefont{and} \bibinfo{author}{\bibfnamefont{B.}~\bibnamefont{Kramer}},
  \bibinfo{journal}{Europhys. Lett.} \textbf{\bibinfo{volume}{24}},
  \bibinfo{pages}{581} (\bibinfo{year}{1993}).

\bibitem[{\citenamefont{Jauregui et~al.}(1996)\citenamefont{Jauregui,
  H\"auseler, Weinmann, and Kramer}}]{Jauregui:prb96}
\bibinfo{author}{\bibfnamefont{K.}~\bibnamefont{Jauregui}},
  \bibinfo{author}{\bibfnamefont{W.}~\bibnamefont{H\"auseler}},
  \bibinfo{author}{\bibfnamefont{D.}~\bibnamefont{Weinmann}}, \bibnamefont{and}
  \bibinfo{author}{\bibfnamefont{B.}~\bibnamefont{Kramer}},
  \bibinfo{journal}{Phys. Rev. B} \textbf{\bibinfo{volume}{53}},
  \bibinfo{pages}{R1713} (\bibinfo{year}{1996}).



\bibitem[{\citenamefont{Szafran et~al.}(2004)\citenamefont{Szafran, Peeters,
  Bednarek, Chwiej, and Adamowski}}]{Szafran:prb04}
\bibinfo{author}{\bibfnamefont{B.}~\bibnamefont{Szafran}},
  \bibinfo{author}{\bibfnamefont{F.~M.} \bibnamefont{Peeters}},
  \bibinfo{author}{\bibfnamefont{S.}~\bibnamefont{Bednarek}},
  \bibinfo{author}{\bibfnamefont{T.}~\bibnamefont{Chwiej}}, \bibnamefont{and}
  \bibinfo{author}{\bibfnamefont{J.}~\bibnamefont{Adamowski}},
  \bibinfo{journal}{Phys. Rev. B} \textbf{\bibinfo{volume}{70}},
  \bibinfo{pages}{035401} (\bibinfo{year}{2004}).

\bibitem[{\citenamefont{H\"ausler}(1996)}]{Hausler:zpb96}
\bibinfo{author}{\bibfnamefont{W.}~\bibnamefont{H\"ausler}},
  \bibinfo{journal}{Z. Phys. B} \textbf{\bibinfo{volume}{99}},
  \bibinfo{pages}{551} (\bibinfo{year}{1996}).

\bibitem[{Kli()}]{Klironomos:cm05}
\bibinfo{note}{A. D. Klironomos, R. R. Ramazashvili, and K. A. Matveev,
  cond-mat/0504118.}

\bibitem[{\citenamefont{Kecke and H\"ausler}(2004)}]{Kecke:prb04}
\bibinfo{author}{\bibfnamefont{L.}~\bibnamefont{Kecke}} \bibnamefont{and}
  \bibinfo{author}{\bibfnamefont{W.}~\bibnamefont{H\"ausler}},
  \bibinfo{journal}{Phys. Rev. B} \textbf{\bibinfo{volume}{69}},
  \bibinfo{pages}{085103} (\bibinfo{year}{2004}).

\bibitem[{\citenamefont{Creffield et~al.}(2001)\citenamefont{Creffield,
  H\"ausler, and MacDonald}}]{Creffield:epl01}
\bibinfo{author}{\bibfnamefont{C.~E.} \bibnamefont{Creffield}},
  \bibinfo{author}{\bibfnamefont{W.}~\bibnamefont{H\"ausler}},
  \bibnamefont{and} \bibinfo{author}{\bibfnamefont{A.~H.}
  \bibnamefont{MacDonald}}, \bibinfo{journal}{Europhys. Lett.}
  \textbf{\bibinfo{volume}{53}}, \bibinfo{pages}{221} (\bibinfo{year}{2001}).

\bibitem[{\citenamefont{Kane et~al.}(1997)\citenamefont{Kane, Balents, and
  Fisher}}]{Kane:prl97}
\bibinfo{author}{\bibfnamefont{C.~L.} \bibnamefont{Kane}},
  \bibinfo{author}{\bibfnamefont{L.}~\bibnamefont{Balents}}, \bibnamefont{and}
  \bibinfo{author}{\bibfnamefont{M.~P.~A.} \bibnamefont{Fisher}},
  \bibinfo{journal}{Phys. Rev. Lett.} \textbf{\bibinfo{volume}{79}},
  \bibinfo{pages}{5086} (\bibinfo{year}{1997}).

\bibitem[{\citenamefont{Voit}(1995)}]{Voit:rpp95}
\bibinfo{author}{\bibfnamefont{J.}~\bibnamefont{Voit}}, \bibinfo{journal}{Rep.
  Prog. Phys.} \textbf{\bibinfo{volume}{58}}, \bibinfo{pages}{977}
  (\bibinfo{year}{1995}).

\bibitem[{\citenamefont{Matveev and Glazman}(1993)}]{Matveev:prl93}
\bibinfo{author}{\bibfnamefont{K.~A.} \bibnamefont{Matveev}} \bibnamefont{and}
  \bibinfo{author}{\bibfnamefont{L.}~\bibnamefont{Glazman}},
  \bibinfo{journal}{Phys. Rev. Lett.} \textbf{\bibinfo{volume}{70}},
  \bibinfo{pages}{990} (\bibinfo{year}{1993}).

\bibitem[{\citenamefont{Cheianov and
  Zvonarev}(2004{\natexlab{a}})}]{cheianov03}
\bibinfo{author}{\bibfnamefont{V.~V.} \bibnamefont{Cheianov}} \bibnamefont{and}
  \bibinfo{author}{\bibfnamefont{M.~B.} \bibnamefont{Zvonarev}},
  \bibinfo{journal}{Phys. Rev. Lett.} \textbf{\bibinfo{volume}{92}},
  \bibinfo{pages}{176401} (\bibinfo{year}{2004}{\natexlab{a}}).

\bibitem[{\citenamefont{Cheianov and
  Zvonarev}(2004{\natexlab{b}})}]{Cheianov04}
\bibinfo{author}{\bibfnamefont{V.~V.} \bibnamefont{Cheianov}} \bibnamefont{and}
  \bibinfo{author}{\bibfnamefont{M.~B.} \bibnamefont{Zvonarev}},
  \bibinfo{journal}{J. Phys. A: Math. Gen.} \textbf{\bibinfo{volume}{37}},
  \bibinfo{pages}{2261} (\bibinfo{year}{2004}{\natexlab{b}}).

\bibitem[{\citenamefont{Fiete and Balents}(2004)}]{Fiete:prl04}
\bibinfo{author}{\bibfnamefont{G.~A.} \bibnamefont{Fiete}} \bibnamefont{and}
  \bibinfo{author}{\bibfnamefont{L.}~\bibnamefont{Balents}},
  \bibinfo{journal}{Phys. Rev. Lett.} \textbf{\bibinfo{volume}{93}},
  \bibinfo{pages}{226401} (\bibinfo{year}{2004}).

\bibitem[{\citenamefont{Haldane}(1981)}]{Haldane:prl81}
\bibinfo{author}{\bibfnamefont{F.}~\bibnamefont{Haldane}},
  \bibinfo{journal}{Phys. Rev. Lett.} \textbf{\bibinfo{volume}{47}},
  \bibinfo{pages}{1840} (\bibinfo{year}{1981}).

\bibitem[{exp()}]{exponent_comment}
\bibinfo{note}{It is worth comparing the exponent of the tunneling density of
  states with earlier results obtain for tunneling into the bulk. Our results
  reduce to the result found in Refs.[\onlinecite{cheianov03}] and
  [\onlinecite{Cheianov04}] in the special case of the infinite $U$ Hubbard
  model where $g=1/2$. In Ref.~[\onlinecite{penc:prl95}] the authors considered
  a situation in which the spin velocity was taken to zero in Luttinger Liquid
  formulas and then the result was Fourier transformed in time, giving an
  exponent -3/8. The calculation of Penc {\em et al.} assumed the electrons
  were always spin-coherent. In essence, they took the limit $T\to 0$ then
  $J\to 0$, the reverse of the limit we consider here and that considered in
  Refs.[\onlinecite{cheianov03}] and [\onlinecite{Cheianov04}].}

\bibitem[{Mue()}]{Mueller:cm04}
\bibinfo{note}{E. J. Mueller, cond-mat/0410773.}

\bibitem[{\citenamefont{Penc et~al.}(1995)\citenamefont{Penc, Mila, and
  Shiba}}]{penc:prl95}
\bibinfo{author}{\bibfnamefont{K.}~\bibnamefont{Penc}},
  \bibinfo{author}{\bibfnamefont{F.}~\bibnamefont{Mila}}, \bibnamefont{and}
  \bibinfo{author}{\bibfnamefont{H.}~\bibnamefont{Shiba}},
  \bibinfo{journal}{Phys. Rev. Lett.} \textbf{\bibinfo{volume}{75}},
  \bibinfo{pages}{894} (\bibinfo{year}{1995}).

\end{thebibliography}

\end{document}